\documentclass{emulateapj}
\usepackage{apjfonts}
\newcommand{\re}{\ensuremath{\rm{R}_e}}
\newcommand{\msol}{\ensuremath{\rm{M}_{\odot}}}
\newcommand{\sm}{\ensuremath{\rm{M}_*}}
\newcommand{\per}{\ensuremath{\!\!\times\!\!}}

\slugcomment{\today}

\shorttitle{Superdense galaxies at low redshift}
\shortauthors{Poggianti et al.}

\begin{document}

%% LaTeX will automatically break titles if they run longer than
%% one line. However, you may use \\ to force a line break if
%% you desire.

\title{Superdense galaxies and the mass-size relation at low redshift}

%% Use \author, \affil, and the \and command to format
%% author and affiliation information.
%% Note that \email has replaced the old \authoremail command
%% from AASTeX v4.0. You can use \email to mark an email address
%% anywhere in the paper, not just in the front matter.
%% As in the title, use \\ to force line breaks.

\author{B.M. Poggianti$^1$, R. Calvi$^{1,2}$, D. Bindoni$^2$, M. D'Onofrio$^2$, A. Moretti$^2$, T. Valentinuzzi$^2$, G. Fasano$^1$, J. Fritz$^3$, G. De Lucia$^{4}$, B. Vulcani$^{1}$, D. Bettoni$^1$, M. Gullieuszik$^1$, A. Omizzolo$^{1}$}
\affil{$^1$INAF-Astronomical Observatory of Padova, Italy,
$^2$Astronomical Department, University of Padova, Italy,
$^3$Sterrenkundig Observatorium Vakgroep Fysica en Sterrenkunde Universiteit Gent, Belgium, $^4$INAF-Astronomical Observatory of Trieste, Italy
}

%% Notice that each of these authors has alternate affiliations, which
%% are identified by the \altaffilmark after each name.  Specify alternate
%% affiliation information with \altaffiltext, with one command per each
%% affiliation.

%% Mark off your abstract in the ``abstract'' environment. In the manuscript
%% style, abstract will output a Received/Accepted line after the
%% title and affiliation information. No date will appear since the author
%% does not have this information. The dates will be filled in by the
%% editorial office after submission.

\begin{abstract}
We search for massive and compact galaxies (superdense galaxies, hereafter
 SDGs) at $z=0.03-0.11$ in the Padova-Millennium Galaxy and Group
Catalogue, a
 spectroscopically complete sample representative of the local
 Universe general field population.  We find that compact galaxies
 with radii and mass densities comparable to high-z massive and
 passive galaxies represent 4.4\% of all galaxies with stellar masses
 above $3 \times 10^{10} M_{\odot}$, yielding a number density of $4.3
 \times 10^{-4} \, \rm h^{3} \, Mpc^{-3}$. Most of them are S0s (70\%)
 or ellipticals (23\%), are red and have intermediate-to-old stellar
 populations, with a median luminosity-weighted age of 5.4 Gyr and
 a median mass-weighted age of 9.2 Gyr.  Their
 velocity dispersions and dynamical masses are consistent with the
 small radii and high stellar mass estimates.  Comparing with the
 WINGS sample of cluster galaxies at similar redshifts, the fraction
 of superdense galaxies is three times smaller in the field than in
 clusters, and cluster SDGs are on average 4 Gyr older than field
 SDGs.  We confirm the existence of a universal trend of smaller radii
 for older luminosity-weighted ages at fixed galaxy mass. As a
 consequence, the median mass-size relation shifts towards smaller
 radii for galaxies with older stars, but the effect is much more
 pronounced in clusters than in the field.  Our results show that,
 on top of the well known dependence of stellar age on galaxy mass,
 the luminosity-weighted age of galaxies depends on galaxy
 compactness at fixed mass, and, for a fixed mass and radius, on environment.  
 This effect needs to be taken into account in order not to overestimate
the evolution of galaxy sizes from high- to low-z. Our
 results and hierarchical simulations suggest that a significant
 fraction of the massive compact galaxies at high-z have evolved into
 compact galaxies in galaxy clusters today.  When stellar age and environmental
 effects are taken into account, the average amount of size evolution of
 individual galaxies between high- and low-z is mild, a factor $\sim
 1.6$.
\end{abstract}

%% Keywords should appear after the \end{abstract} command. The uncommented
%% example has been keyed in ApJ style. See the instructions to authors
%% for the journal to which you are submitting your paper to determine
%% what keyword punctuation is appropriate.


%% From the front matter, we move on to the body of the paper.
%% In the first two sections, notice the use of the natbib \citep
%% and \citet commands to identify citations.  The citations are
%% tied to the reference list via symbolic KEYs. The KEY corresponds
%% to the KEY in the \bibitem in the reference list below. We have
%% chosen the first three characters of the first author's name plus
%% the last two numeral of the year of publication as our KEY for
%% each reference.

\section{Introduction}

%Understanding how massive galaxies managed to assemble their mass
%already at high redshift is one of the biggest challenges in cosmology
%today. 

Recent observations have found that the sizes of high-z quiescent and
massive galaxies at $z \sim 1-2.5$ are on average much smaller than
those of high Sersic index ($n \geq 2.5$) galaxies in the local
Universe (Daddi et al. 2005, Trujillo et al. 2006, Toft et al. 2007,
Zirm et al. 2007, Buitrago et al. 2008, Cimatti et al. 2008, van
Dokkum et al. 2008, van der Wel et al. 2008, Saracco et al. 2009,
Cassata et al. 2011, Damjanov et al. 2011). Recent tantalizing
observations from CANDELS may have identified the $z>2$ compact,
star-forming progenitors of the passive $z=1-2$ population (Barro
et al. 2012). At a given stellar mass, $z=1-2$ passive galaxies
exhibit a wide range of sizes, from those with radii comparable to
nearby galaxies to ultra-compact galaxies up to six times smaller
(Saracco et al. 2009, Mancini et al. 2010, Cassata et al. 2011).
Searches for superdense galaxies in the Sloan at low-z have found very
few such galaxies in the general field (Trujillo et al. 2009, Taylor
et al. 2010).

These results suggest a strong evolution in galaxy size between $z=2$
and $z=0$.  The galaxy size growth appears to be
stronger for more massive galaxies (Ryan et al. 2012, Huertas-Company
et al. 2012), and at least at $z<1$ dependent on morphological
type (Huertas-Company et al. 2012).

Several mechanisms have been proposed to explain these
observations. One such mechanism is 
expansion for mass loss, due to an active
galactic nucleus (Fan et al. 2008) or stellar winds (Damjanov et
al. 2009), though Ragone-Figueroa \& Granato (2011) have found that
these are insufficient to explain the observed evolution. 
Major mergers are another possible process at work, but the major merger
rate seems to be insufficient to explain the observed evolution.
More frequent minor mergers are currently considered the most likely
explanation for the evolution of galaxy sizes, since they
reduce the effective stellar densities, strongly
increasing the galaxy size while leaving the mass almost unaffected
(Khochfar \& Silk 2006, Naab et al. 2009, Hopkins et al. 2009,
Bezanson et al. 2009, Hilz et al. 2012, Oser et al. 2012, see also
Trujillo et al. 2011).  
However, several studies
%yield a size evolution of a factor 1.4-1.8, they
struggle to reconcile the minor merger scenario 
with the large amount of evolution required when comparing at
face value high- and low-z galaxy populations  
(Nipoti et al. 2009, 2012, Saracco et al. 2011,
Newman et al. 2012, Oogi \& Habe 2012).

Two new aspects that have been recently highlighted may lead to a
modification of the current paradigm of size evolution. The first one
has been the detection of a significant fraction of superdense
galaxies with radii and masses comparable to high-z galaxies in
nearby galaxy clusters (Valentinuzzi et al. 2010a, hereafter V10) and in clusters at
intermediate redshift (Valentinuzzi et al. 2010b). This may imply a
strong environmental dependence of the distribution of massive and
compact galaxies in the local Universe, but no study so far has 
carried out a detailed comparison between clusters and field at low-z.

The second aspect concerns the effects of galaxy selection in the
high-z samples. Several studies have found that smaller galaxies have
older stellar populations. As a result, by selecting galaxies that are
already passive, high-z studies might be preferentially selecting the
most compact ones (Valentinuzzi et al. 2010, Saracco et al. 2009,
2011, Cassata et al. 2011, Szomoru et al. 2011).  The existence of a
trend of size with age at fixed mass is however still debated
(e.g. cf. Valentinuzzi et al. 2010 and Trujillo et al. 2012).

In this paper we search for superdense galaxies and analyze the
mass-size relation of galaxies in a new sample representative of the
general field population in the local Universe, the Padova-Millennium
Galaxy and Group Catalogue (PM2GC, Calvi et al. 2011).  This is the
first non-Sloan-based work of this kind in the field at low-z, on a much
smaller sky area than Sloan, but with a better quality imaging and higher
spectroscopic completeness.

We investigate the incidence and properties of superdense galaxies as
a function of environment, comparing the general field with clusters,
and a range of finer environments including isolated galaxies and
groups.  We also focus on the dependence of the mass-size relation on
galaxy ages and environment, attempting to quantify the importance of
selection effects when inferring the size evolution of individual
galaxies.

After presenting our dataset (\S2), in \S3 we show our results for the
fraction, number density and properties of superdense galaxies in the
field, comparing with clusters in \S4.  The mass-size relation for
different types of galaxies, different stellar population ages and as
a function of environment is discussed in \S5. We conclude by
comparing with previous works (\S6) and summarizing our results (\S7).
We release our surface photometry catalog describing it in the
Appendix.

Throughout this paper we will use the cosmology ($H_0$, ${\Omega}_m$,
${\Omega}_{\lambda}$) = (70,0.3,0.7).  All the masses in this paper
have been scaled to a Kroupa (2001) IMF (0.1-100 $M_{\odot}$).

\section{The dataset}

We draw our galaxy sample from the Padova-Millennium Galaxy and Group
Catalogue (hereafter, PM2GC, Calvi et al. 2011) consisting of a
spectroscopically complete sample of galaxies at $0.03 \leq z \leq
0.11$ brighter than $M_B < -18.7$. This sample is sourced from the
Millennium Galaxy Catalogue (MGC, Liske et al. 2003, Driver et
al. 2005), a B-band contiguous equatorial survey of $\sim 38 \, deg^2$
complemented by a 96\% spectroscopically complete survey down to B=20.
%The image quality and the spectroscopic completeness are superior to
%Sloan (e.g. only 86\% of the PM2GC galaxies have SDSS spectra), and these
%qualities make it an interesting dataset to study galaxy sizes in
%a complete sample of galaxies at low-z.
The image quality and the spectroscopic completeness are superior to
Sloan, and these qualities make it an interesting dataset to study
galaxy sizes in a complete sample of galaxies at low-z. In particular,
the MGC is based on INT data (2.5m telescope) obtained with a median
seeing of 1.3'' and at least 750 sec of exposure, with a pixel scale
of 0.333''/pixel, while the Sloan (again 2.5m telescope) has a
median seeing of 1.5'' in $g$ (the closest band to our data), an
exposure time of only 54.1 sec and 0.396''/pixel. As for
completeness, 14\% of all galaxies and 25\% of SDGs in our
spectroscopic PM2GC sample do not have an SDSS spectrum.

The whole PM2GC sample consists of 3210 galaxies and is representative
of the general field population. Within the PM2GC,
Calvi et al. (2011) identified 176 galaxy groups with at least three
members brighter than $M_B < -18.7$ using a Friends-of-Friends group-finding
algorithm.  A galaxy is a group member if its spectroscopic redshift
lies within $\pm 3\sigma$ from the median group redshift and if it is
located within a projected distance of $\leq 1.5 R_{200}$\footnote{$R_{200}$ 
is defined as the radius of a sphere whose interior mean density is 200 times
the critical density of the Universe, and it approximates the cluster
virial radius.} from the
group geometrical centre.  Galaxies which have no neighbour or solely
one with a projected mutual distance $\leq 0.5 \rm \, h^{-1} \, Mpc$
and a redshift within 1500$\rm \, km \, s^{-1}$ are considered
``single'' galaxies or ``binary-system'' galaxies, respectively.

The PM2GC sample is complete for masses $M_{\star} \geq 1.8 \times
10^{10} M_{\odot}$ ($log M_{\star} \geq 10.25$), corresponding to the
mass of the faintest and reddest galaxy ($M_B =-18.7$, B-V = 0.9) at
our redshift upper limit (z = 0.1), as described in Calvi et
al. (2012).  In this paper, however, we consider only those 1057
galaxies with $M_{\star} \geq 3 \times 10^{10} M_{\odot}$ ($log
M_{\star} \geq 10.48$) to be consistent with the cluster work from
Valentinuzzi et al. (2010).  Surface profile fits were obtained for
94\% of these galaxies as outlined below.  Hereafter, we will use the
mass-limited sample with radii estimates consisting of 995 galaxies at
z=0.03-0.11.

\subsection{Galaxy radii and Sersic indices}

Effective-radii, axial ratios   and Sersic indexes are  measured
on MGC  B-band   images  with GASPHOT  (Pignatelli \& Fasano et al. 2006,
Bindoni et al. in prep.),  an
automated tool  which performs  a simultaneous  fit  of the  major and
minor  axis light  growth curves    with  a 2D flattened   Sersic-law,
convolved by the appropriate,  space-varying PSF.  In this way GASPHOT
exploits the robustness of  the 1D fitting  technique, keeping at  the
same time the capability (typical of the 2D  approach) of dealing with
PSF convolution in the innermost regions.

GASPHOT has  proved to be very  efficient in  recovering the best fitting
parameters, and  to give the  appropriate weight to the external parts
of the  galaxies, where PSF effects  are  negligible. Indeed we tested
GASPHOT on  more than  15,000  simulated and real  galaxies, obtaining
robust  upper  limits  for the  errors of   the  global parameters  of
galaxies,    even  for    non-Sersic   profiles  and   blended objects
(Pignatelli \& Fasano 2006). The $\Delta R_e/R_e$ error (median difference
between retrieved and input value) is
-0.004$\pm$0.003 and -0.030$\pm$0.024 for simulated $n=1$ and $n=4$
galaxies, respectively, and -0.057$\pm$0.101 for real galaxies.
GASPHOT was also tested against the widely used tools GALFIT (Peng et
al. 2002) and GIM2D (Marleau \& Simard 1998). It was found
(Pignatelli \& Fasano 2006) that the performances of these tools are
quite similar for large and regular simulated galaxies, while GASPHOT
has proved to be more robust for real galaxies with some kind of
irregularity, which is a crucial feature when dealing with
blind surface photometry of huge galaxy samples.

The  GASPHOT  output effective radius  \re\  value is
calculated  along the major-axis, and for the purposes of this
paper is circularized with
the usual formula:
\begin{equation}
\re^{(circ)}=\re^{(major)}\cdot\sqrt{b/a}
\end{equation}  
where $a$ and $b$ are the major- and minor-axis of the best-fit model,
respectively.

\begin{figure*}
\centering
\centerline{\includegraphics[scale=0.8]{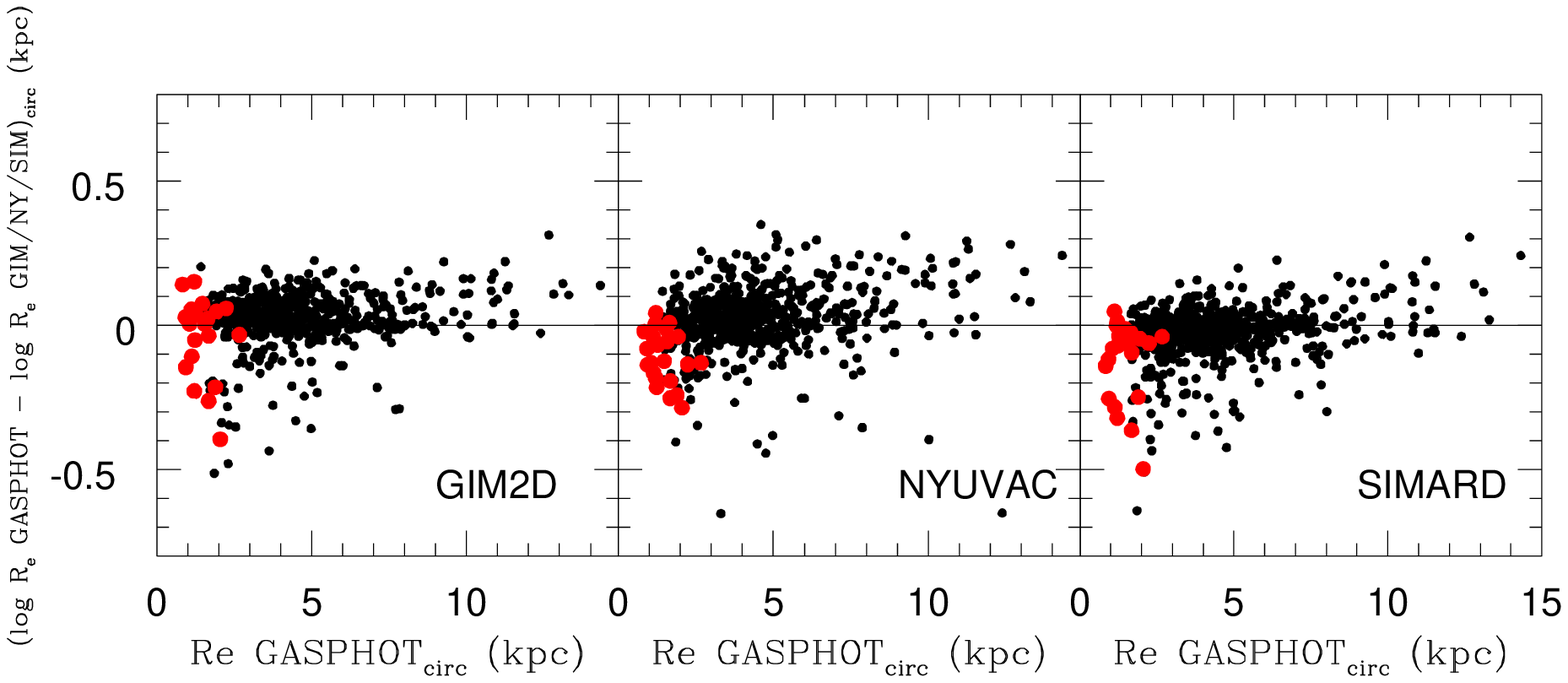}\vspace{-9cm}}
\centerline{\includegraphics[scale=0.8]{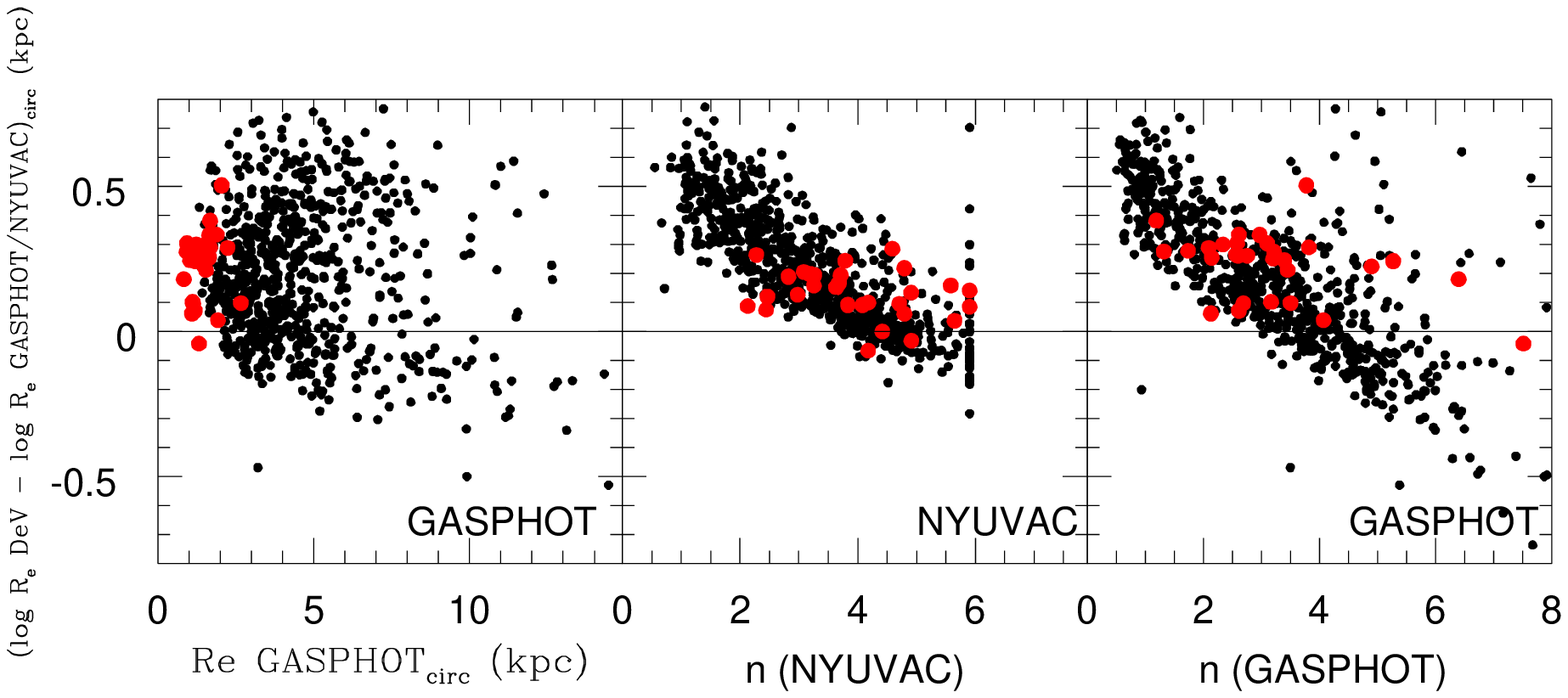}}
%\includegraphics[scale=0.6]{papre.ps}
%\vspace{-10cm}
%\includegraphics[scale=0.8]{papre2new.ps}
\caption{Top. Comparison between circularized GASPHOT radii and GIM2D/NYUVAC/SIMARD radii all in kpc.
Bottom. Comparison between de Vaucoulers DR7 radii (g band)
and GASPHOT/NYUVAC radii as a function of GASPHOT radius and
Sersic index. SDGs are red circles.
\label{fig:re}}
\end{figure*}

Besides our own size measurements, there are four other independent
size estimates for PM2GC galaxies: GIM2D Sersic fits from the
Millennium Galaxy Catalogue data (Allen et al. 2006), New York University
Value Added Catalogue (NYUVAC) radii (Blanton et al. 2005), GIM2D
Sersic fits from Sloan DR7 data (Simard et al. 2011), and DR7/SDSS de
Vaucouleurs sizes accessed via the Catalog Archive Server (Abazajian
et al. 2009, Thakar et al. 2008), the last three being based
on Sloan $g$-band data.

Figure~\ref{fig:re} shows the comparison between GASPHOT and
GIM2D(MGC)+NYUVAC+SIMARD size estimates for the 618 galaxies in common
among the four samples.  The agreement is generally good, with a
tendency for GASPHOT radii to be larger than the others.  The median
difference between PM2GC radii and GIM2D/NYUVAC/SIMARD radii is
0.03$\pm$0.04/0.03$\pm$0.06/-0.01$\pm$0.04dex for all galaxies and
0.01$\pm$0.08/-0.12$\pm0.07$/-0.06$\pm$0.11dex for superdense galaxies
(SDGs) as defined in \S3.  Therefore, for SDGs the Sloan-based
estimates tend to be slightly larger than the MGC-based estimates, by
a factor 1.3 and 1.1 for the NYUVAC and the Simard catalogues,
respectively. The effect is clearly very small, especially when using
GIM2D on Sloan data.

Since there are previous low-z studies that have used
de Vaucoulers DR7 radii (Taylor et al. 2010, Trujillo et al. 2011),
in Fig.~\ref{fig:re} we also show the comparison between DR7 size estimates
obtained for a de Vaucoulers (n=4) fit and both GASPHOT and NYUVAC
radii (764 in common among the three samples). 
The NYUVAC estimates have been obtained from 
the same data (Sloan g-band imaging)
of the de Vaucoulers radii. As could be expected, the agreement
is good for $n \sim 4$, but de Vaucoulers radii are much larger 
for lower $n$ values, where most of the galaxies and, as we will see,
most of our SDGs are found.

\subsection{Galaxy stellar masses}

\begin{figure}
\centering
\includegraphics[scale=0.55]{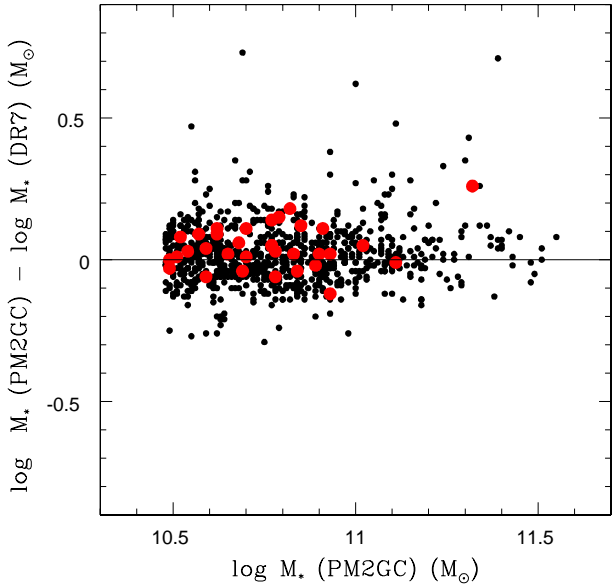}
\caption{Comparison between the galaxy stellar masses used in this paper
(PM2GC) and DR7 masses. Red galaxies are SDGs, black are galaxies in
the mass-limited sample.
\label{fig:masses}}
\end{figure}

Galaxy stellar mass estimates were derived by Calvi et al. (2011)
using the Bell \& de Jong (2001) relation which
correlates the stellar mass-to-light ratio with the optical colours
of the integrated stellar population, using the B-band photometry taken
from the MGC, and the rest-frame B-V color computed from the Sloan
$g-r$ color corrected for Galactic extinction
%, we used the Bell \& De Jong
%(2001) coefficient for a solar metallicity Bruzual \& Charlot model
(see Calvi et al. 2011 for details). The typical uncertainty on
stellar mass estimates is 0.2-0.3dex.

Figure~\ref{fig:masses} shows that there is no systematic difference
between our mass estimates and 
DR7\footnote{http:\/\/www.mpa-garching.mpg.de\/SDSS\/DR7\/Data\/stellar mass.html} 
ones, and the scatter is $<0.1$dex.
The median mass difference ($log M_{\star} (PM2GC) - log M_{\star}$
(DR7)) is -0.01$\pm$0.07dex and 0.02$\pm$0.1dex for all and SDG
galaxies, respectively.

\subsection{Morphologies}

To morphologically classify galaxies, we used MORPHOT, an automatic tool
designed to reproduce as closely as possible the visual classifications
(Fasano et al. 2012). MORPHOT adds to the classical CAS (concentration/
asymmetry/clumpiness) parameters a set of additional indicators
derived from digital imaging of galaxies and
has been proved to give an uncertainty only slightly larger than
the eyeball estimates (Fasano
et al. 2011). MORPHOT was applied to the B-band MGC images to
identify ellipticals, lenticulars (S0s) and later-type galaxies.
These are the three broad morphological classes we will use in this paper (see 
Calvi et al. 2012 for details).

It is well known that separating ellipticals from S0s could be sometimes
challenging even for human experienced classifiers. The challenge is
obviously enhanced when dealing with compact objects like SDGs. That
being stated, the MORPHOT ability of recognizing visually classified S0
galaxies on MGC-like imaging has been shown to be of the order of
$\sim 80$\% (see Figure 11 in Fasano et al. 2012; see also Figures 6 and 16
therein), which is comparable to the level of agreement reached
by different experienced visual classifiers.

\subsection{Stellar population ages}

The MGC spectroscopic database of PM2GC galaxies consists of SDSS,
2dFGRS and MGCz spectra (Driver et al. 2005), the latter taken with the same
instrument and setup of the 2dFGRS. Choosing always the highest
quality spectrum available, we preferred to use SDSS spectra when
possible (858 cases, 86\% of the sample), alternatively 2dFGRS spectra
when available (93 cases) and MGCz spectra in the remaining
cases. SDSS spectra are flux-calibrated, while we performed a relative
flux calibration for all 2dF spectra using the response curve provided
by the 2dF team (http:\/\/www.mso.anu.edu.au\/2dFGRS\/) 
and a refined, second order calibration using a
curve derived by the ratio between the 2dF spectra and the
corresponding SDSS spectra for objects with both spectra available, as
described in Cava et al. (2009).

To derive the stellar populations ages, we fitted the spectra with the
spectro-photometric model fully described in Fritz et al. (2007).  All
the main spectro-photometric features (such as the continuum flux and
shape, and the equivalent widths of emission and absorption lines) are
reproduced by summing the theoretical spectra of Simple Stellar
Population (SSP) of 12 different ages (from $3\per 10^6$ to $\sim
14\per10^9$ years).

Dust extinction is allowed to vary as a function of SSP age, in a
screen uniformly distributed in front of the SSP stars.  The Galactic
extinction law follows the Cardelli et al. (1989) scheme, with $R_V=3.1$.  As
explained in detail in Fritz et al. (2007), for the fit we use a fixed
metallicity, exploring three values: Z=0.004, Z=0.02 and Z=0.05. The
adopted star formation histories refer to the model
with the metallicity value that provides the lowest $\chi^2$.
%The lowest  $\chi^2$ for the great majority of our  super-dense  galaxies
%yields either solar or supersolar metallicities.

SSP spectra are built using Padova evolutionary tracks and the
observed MILES spectral library (Sanchez-Blazquez et al. 2006) for ages
older than $10^9 \rm \, yr$, complemented by the Jacoby et al. (1984)
library for young SSPs, and in the UV and infrared by means of the
Kurucz theoretical library.  Nebular emission is included,
modeled with values that are typical of {\sc Hii} regions: this
significantly affects spectra of SSPs younger than $\sim2\per10^7$
years.

From our spectral analysis, it is possible to derive an estimate of
the average age of the stars in a galaxy. Following the definition of
Cid-Fernandes et al. (2005), we compute the luminosity--weighted (LW) age by
weighting the age of each SSP composing the integrated spectrum with
its bolometric flux. This provides an estimate of the average age of
the stars weighted by the light we actually observe.  A mass--weighted (MW)
age is computed in a similar way: each SSP age is weighted with its
mass value.  The mass--weighted age is the ``true'' average age of the
galaxy's stars. By construction,  the MW age is always older than the LW
age.

\section{Results: Superdense galaxies in the field at low redshift}

\begin{figure*}
\centering
\includegraphics[scale=0.6]{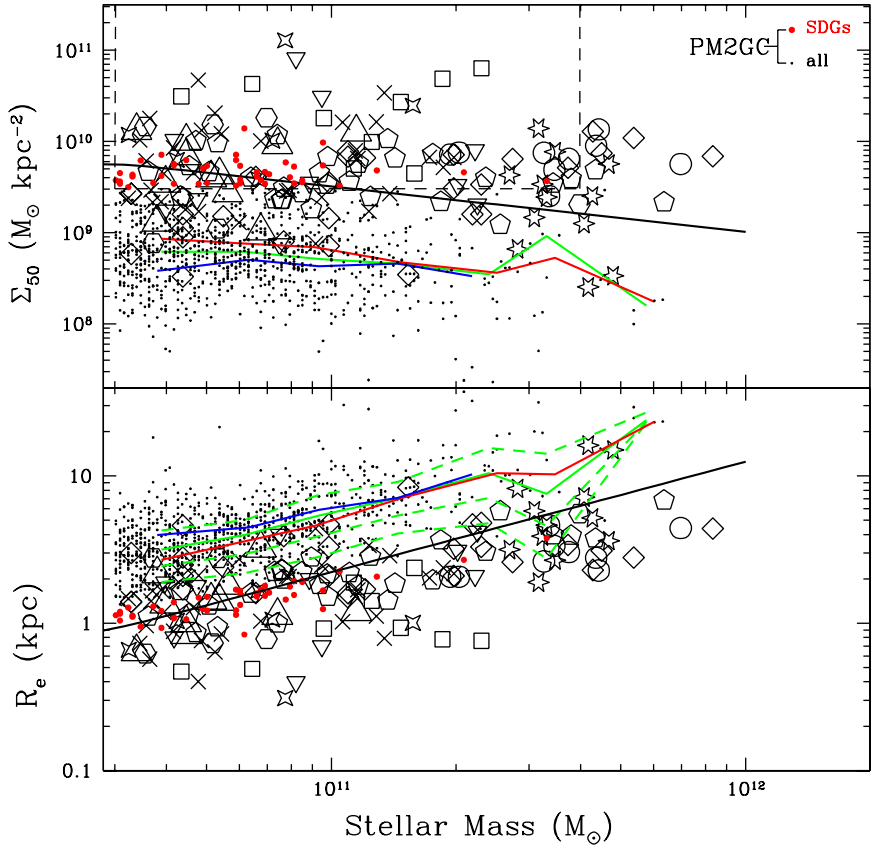}
\caption{The circularized effective radius $R_e$ and the mass-density
inside $R_e$ as a function of stellar mass for all PM2GC galaxies with
$M_*\geq10^{10.48}\msol$.  
The region  corresponding to our SDGs definition  is delimited  by the
dashed  lines in the top  panel.   The corresponding larger red  
dots mark  the SDGs.    The black solid  line  is
delimiting the ATLAS3D zone of avoidance (see text).
The green (with dotted $1\sigma$ and $2\sigma$ lines), red 
and blue lines are the PM2GC median relations
for all galaxies and for early and late-type galaxies, respectively.
Open symbols  are galaxies   from high-z  studies, see  text   for
references.
\label{fig:all}}
\end{figure*}

In the bottom  panel  of Fig.\ref{fig:all}, we  plot  the circularized
effective   radius  \re\    as a    function   of  stellar  mass   for
PM2GC galaxies with stellar masses
$\sm\geq10^{10.48}\msol$.   In the upper-panel,   we  plot the  mean mass
surface density inside \re:
\begin{equation}
\Sigma_{50} = \frac{0.5\sm}{\pi\re^{2}}
\end{equation} 
as   usually  defined   by     other    authors (see, among others, Cimatti
et al. 2008 and van der Wel et al. 2008).
The plots include the median PM2GC mass-size relation for all (green,
with 1$\sigma$ and 2$\sigma$ dashed lines), early-type (ellipticals + S0s, red
line) and late-type (blue line) galaxies. The binned values of these mass-size
relations are given in Table~1.

High-z literature data at $1<z<2.5$ are plotted as large empty
symbols: HUDF (Daddi et al. 2005, exagons), 
MUNICS (Trujillo et al. 2006, circles), MUSYC
(van Dokkum et al. 2008, squares), Saracco et al. (2009) (diamonds), 
GMASS (Cimatti et
al. 2008, triangles), van der Wel et al. (2008) (pentagons), Damjanov et al. 
(2009) (reversed triangles), Mancini
et al. (2010) (stars), Cassata et al. (2011) (crosses $1.2<z<2$, 
starred crosses $z>2$). 
All data have been converted to a Kroupa IMF (Kroupa 2001)
and Maraston (2005) models as done in V10.
The data, methods of analysis,
galaxy mass ranges and selection criteria differ from one study to
another, but all these works, with the exception of van der Wel et
al. (2008) that have used a visual morphological classification, have
selected galaxies to be already passively evolving with old stellar
populations at the redshift they are observed, based on SED spectral
fitting, absence of emission lines, line index age dating and other
spectrophotometric techniques.

Following V10, we define SDGs trying to match as much as possible the
position of most high-z SDG data in Fig.~\ref{fig:all} applying the following
criteria:

\begin{eqnarray}
 3\per10^{10}\msol & \leq & \sm  \leq  4\per10^{11}\msol \\
 \Sigma_{50} & \geq & 3\per10^{9}\msol kpc^{-2}
\end{eqnarray}

SDG candidates complying these criteria
have been individually inspected, to check their size
and mass estimates. A merging system has been excluded from the analysis. 
When the GASPHOT and GIM2D estimates significantly
disagreed, a galaxy-by-galaxy analysis performed with {\it iraf} 
was used to decide
which radius was the most accurate.
Most of these cases are galaxies with a nearby galaxy or star.
%The AIAP size was adopted if larger than the GASPHOT size. ({\bf Gianni,
%explain why
%in these cases AIAP is better}).
There were three galaxies for which the DR7 mass estimate agreed
with the mass obtained using SDSS CAS magnitudes but disagreed
with the PM2GC mass. For these, the SDSS CAS mass was adopted.
This left a total sample of 44 SDGs.

The fraction of SDGs among PM2GC galaxies with masses 
$3 \times 10^{10} < M_{\star} < 4 \times 10^{11} M_{\odot}$ 
is $4.4\pm 0.7$\%. %44/988
The corresponding number density of SDGs within the PM2GC
volume is $4.3 \times 10^{-4} \, \rm
h^{3} \, Mpc^{-3}$.
%(seems good agreement with Cassata, check their fig.8).

We note that if we used only literature information the fraction of
SDGs would be only slightly lower, but still comparable within the
statistical error, 3.7$\pm 0.7$\% using GIM2D radii and DR7 masses and
3.3$\pm 0.7$\% for NYUVAC radii and DR7 masses.

We also note that the fraction of PM2GC galaxies and PM2GC SDGs with a
size measurement from the NYUVAC are 85\% and 73\%, and the fractions
with DR7 masses are 86\% and 75\%, respectively. This supports the
notion that the SDSS sample might be indeed slightly biased against
compact massive galaxies (less complete for compact than
non-compact galaxies), as discussed in detail 
by other authors (e.g. Taylor et al. 2010).

It can be seen from  Fig.\ref{fig:all} that our SDG definition
corresponds to galaxies with sizes more than 2$\sigma$ smaller than the
median mass-size relation. The region at radii below and
densities above the solid straight line is the ''zone of avoidance''
where galaxies are rare in the ATLAS3D sample of 260 nearby early-type
galaxies, a complete and statistically representative sample of the nearby 
early-type population (Cappellari et al. 2011). Our SDGs are truly
exceptionally dense objects, and outliers in the local mass-radius
and mass-density relations.

\begin{figure*}
\centering
\includegraphics[scale=0.6]{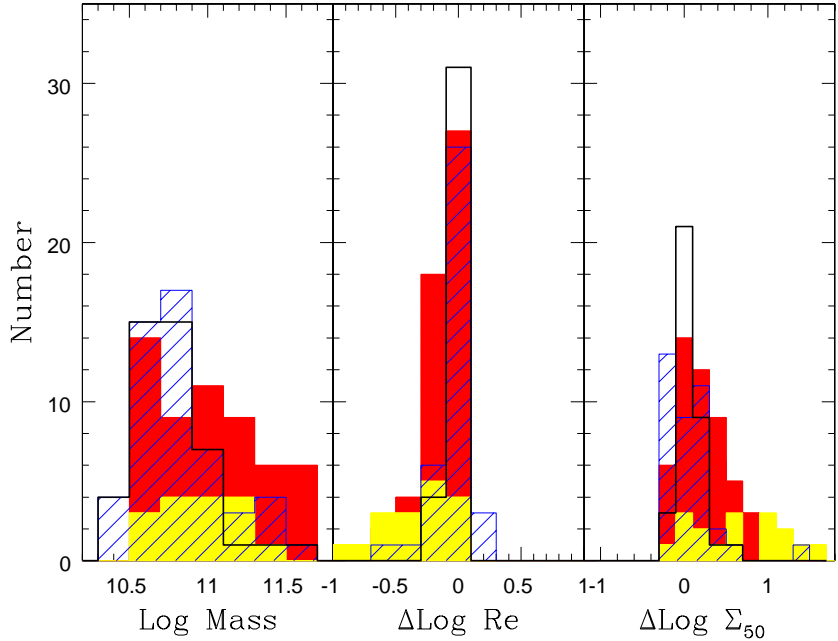}
\caption{Distribution of stellar mass, 
offset of effective radius from the high-z median mass-size relation
and offset of density $\Sigma_{50}$ from the high-z median mass-density
relation. Symbols: PM2GC SDGs (black histogram),
WINGS SDGs (blue shaded histogram) and high-z samples at $z<2$ (red histogram) 
and $>2$ (yellow). Only high-z galaxies fulfilling SDG criteria (with 
$ 3\per10^{10}\msol \leq M_{\star}  \leq  4\per10^{11}\msol$ and
$ \Sigma_{50} \geq  3\per10^{9}\msol kpc^{-2}$) are included
in this plot. Note that the median mass-size and mass-density relations
used here include {\it all} galaxies at high-z, and this is the reason
why the red+yellow histograms are not centered on 0.
\label{fig:distri2}}
\end{figure*}

In Fig.~\ref{fig:distri2} we compare the distributions of galaxy mass,
radius and density values of the PM2GC SDG sample with those of the
high-z SDG data, together with the cluster low-z WINGS sample that
will be discussed later. We do this to check that in the local
Universe we are selecting galaxies with a "degree of compactness" that
is comparable to that of high-z galaxies.  For masses, we plot the
simple distributions, while for radii and densities we plot the offset
with respect to the high-z median mass-radius and mass-density
relations obtained using all the high-z samples considered in this
paper.

The PM2GC, WINGS and high-z distributions largely overlap (cf.
Fig.7 in Taylor et al. 2010).  However, the low-z and high-z
distributions do not coincide, and they are not expected to,
simply because the low-mass datasets are mass-limited galaxy samples,
while the high-z histograms are the results of a collection of
different surveys, with different mass ranges, selection criteria
etc. As a result, the mass distributions at high- and low-z
differ, with the high-z histogram being more populated at high masses
than the local samples, and the consequent differences are visible
also in the radii and density panels.  Nevertheless, the large overlap
of the $\Delta Log R_e$ and $\Delta Log \Sigma_{50}$ distributions of
low-z and high-z SDGs shows that we are indeed comparing local and
distant samples of galaxies with a large overlap in degree of
compactness and density.

We note however that $z>2$ datapoints (yellow histograms in
Fig.~\ref{fig:distri2}) have more prominent low-radii and
high-densities tails with respect to the $z<2$ distributions. This is
mainly due to the van Dokkum et al. (2008) sample, while Cassata et
al. (2011) datapoints resemble the $z<2$ distribution. It is unclear
whether the fact that $z>2$ galaxies seem to be extreme in their
compactness is due to rapid evolution between $z=2$ and $1$ (Cimatti
et al. 2012) or selection effects (Saracco et al. 2010), but
observationally the difference with $z$ appears quite strikingly and
is worth further investigation.

\begin{table}
\begin{center}
\begin{tabular}{lcc}
%\hline
%\hline
\hline
 $\log_{10}(\sm/\msol)$ & \multicolumn{2}{c}{$\log_{10}(\re/\rm{kpc})$} \\
\hline
 & Late Types & \\
% & \multicolumn{2}{c}{Late Type galaxies} \\
 &  $T>0$ & $ n < 2.5$ \\ 
\hline
$10.58_{-0.04}^{+0.06}$  &  $0.60_{-0.09}^{+0.12}$  & $0.54_{-0.10}^{+0.13}$ \\
$10.80_{-0.05}^{+0.04}$  &  $0.65_{-0.10}^{+0.09}$  & $0.63_{-0.12}^{+0.10}$ \\
$10.97_{-0.04}^{+0.06}$  &  $0.77_{-0.10}^{+0.12}$  & $0.74_{-0.12}^{+0.10}$ \\
$11.16_{-0.04}^{+0.05}$  &  $0.85_{-0.10}^{+0.10}$  & $0.78_{-0.10}^{+0.06}$ \\
$11.34_{-0.03}^{+0.05}$  &  $1.01_{-0.19}^{+0.11}$  & $0.79_{-0.11}^{+0.05}$ \\
\hline
 & Early Types & \\
% & \multicolumn{2}{c}{Early Type galaxies}\\
 &  $\rm T \leq 0$ & $n \geq 2.5$ \\  
\hline
$10.59_{-0.05}^{+0.04}$  &  $0.43_{-0.13}^{+0.09}$  &  $0.46_{-0.12}^{+0.11}$ \\
$10.78_{-0.04}^{+0.06}$  &  $0.55_{-0.15}^{+0.11}$  &  $0.58_{-0.13}^{+0.11}$ \\
$10.96_{-0.03}^{+0.05}$  &  $0.66_{-0.12}^{+0.12}$  &  $0.72_{-0.14}^{+0.16}$ \\
$11.17_{-0.05}^{+0.04}$  &  $0.85_{-0.14}^{+0.13}$  &  $0.89_{-0.13}^{+0.11}$ \\
$11.40_{-0.08}^{+0.03}$  &  $1.02_{-0.25}^{+0.22}$  &  $1.04_{-0.16}^{+0.20}$ \\
$11.54_{-0.03}^{+0.12}$  &  $1.01_{-0.36}^{+0.23}$  &  $0.89_{-0.29}^{+0.29}$ \\
$11.78_{-0.05}^{+0.02}$  &  $1.37_{-0.01}^{+0.10}$  &  $1.38_{-0.01}^{+0.05}$ \\
\hline
 &  All galaxies & \\ 
\hline
$10.59_{-0.05}^{+0.04}$  &  $0.50_{-0.11}^{+0.13}$  & \\
$10.79_{-0.05}^{+0.05}$  &  $0.60_{-0.13}^{+0.10}$  & \\
$10.97_{-0.04}^{+0.05}$  &  $0.73_{-0.14}^{+0.14}$  & \\
$11.16_{-0.04}^{+0.05}$  &  $0.85_{-0.12}^{+0.11}$  & \\
$11.38_{-0.06}^{+0.03}$  &  $1.02_{-0.17}^{+0.17}$  & \\
$11.52_{-0.01}^{+0.08}$  &  $0.88_{-0.22}^{+0.27}$  & \\
$11.76_{-0.03}^{+0.03}$  &  $1.38_{-0.01}^{+0.05}$  & \\
\hline 
\end{tabular}
\caption{PM2GC mass-radius relations. The values are logarithm of
the median estimates, errors are the lower  and upper quartiles of the
distributions.\label{tab:wlocal}}
\end{center}
\end{table}

\subsection{The properties of superdense field galaxies at low-z}

\begin{figure}
%\centering
\centerline{\includegraphics[scale=1.2]{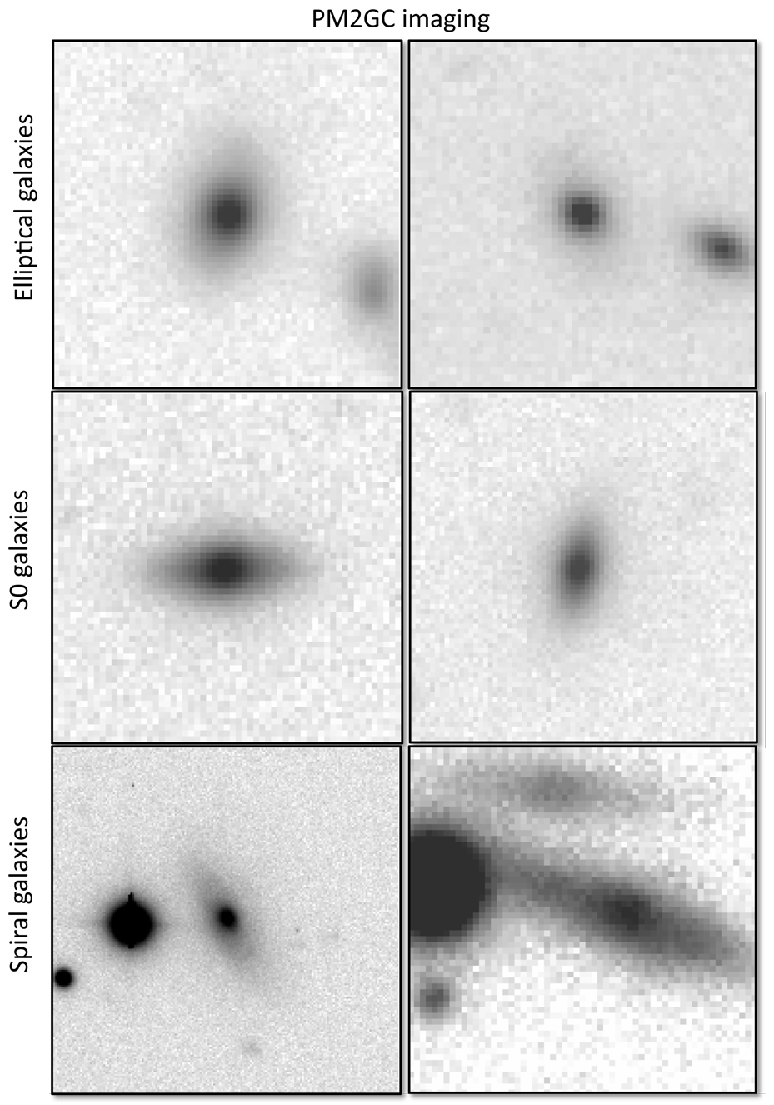}}
% provare anche stamps2.eps
\caption{Example postage stamps of PM2GC elliptical, S0 and late-type SDGs.
\label{fig:gianni}}
\end{figure}

23$\pm7$\% of our SDGs 
are ellipticals, 70$\pm13$\% are S0s and only 7$\pm4$\% 
are later-type galaxies. Figure~\ref{fig:gianni} shows some
example postage stamps of PM2GC elliptical, S0 and late-type SDGs.
In the whole mass-limited sample, 29\% are ellipticals, 30\% are S0s
and 41\% are late-types.\footnote{Comparing with the Sloan Bayesian 
automated morphological classifications
of Huertas-Company et al. (2011), our morphological classifications
agree with theirs ($P_{Ttype} > 0.45 $, our S0s versus their S0+early-spirals) 
in 79\% of the cases for the 33 SDGs in common.}

\begin{figure*}
%\centering
\centerline{\includegraphics[scale=0.6]{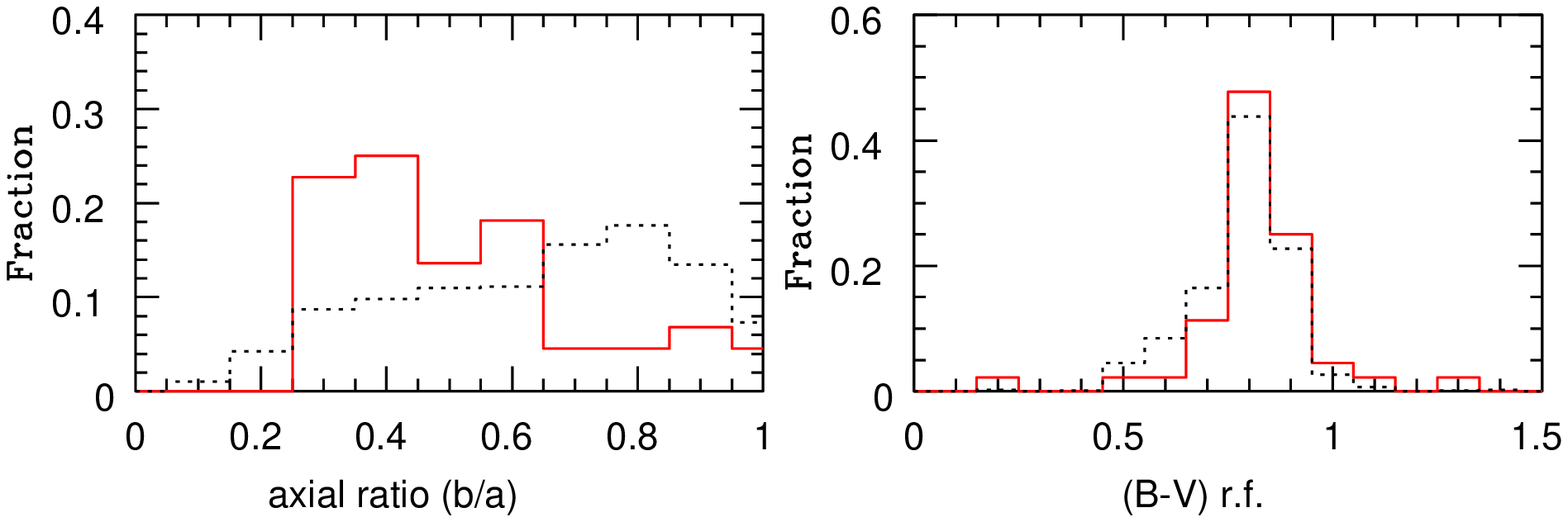}\vspace{-0.5cm}}
\centerline{\includegraphics[scale=0.6]{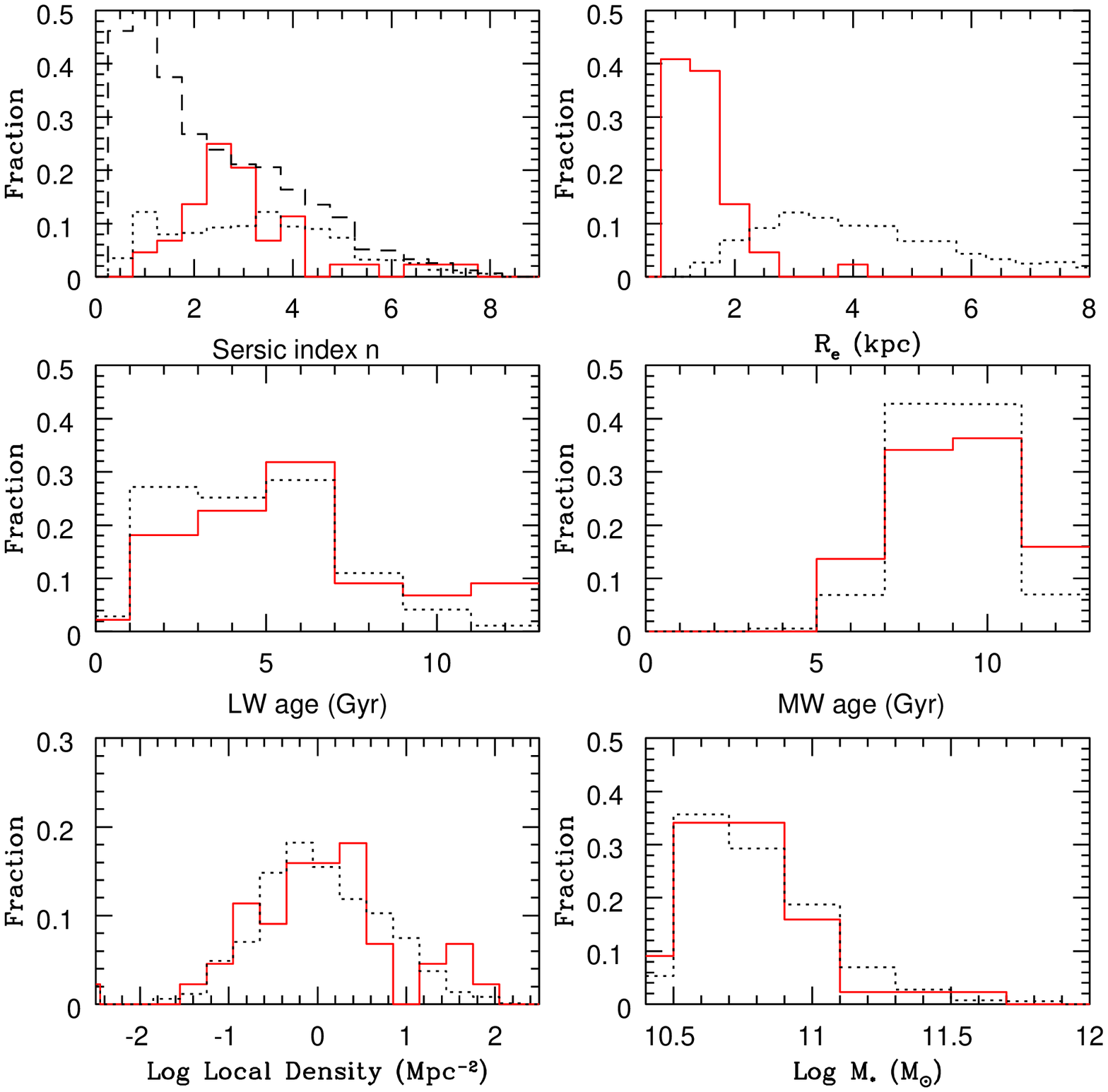}}
\caption{Distribution of different quantities of interest for PM2GC SDGs
(red solid histogram) and non-SDGs (black dotted): axial ratio, Sersic index,
circularized effective radius, LW age, MW age, local density and stellar mass. 
The dashed black histogram
in the Sersic index panel is the $n$ distribution for the absolute 
magnitude-limited sample (see text).
\label{fig:allprops}}
\end{figure*}

In Fig.~\ref{fig:allprops} we present the distributions of the most
relevant quantities describing our SDG sample: axial ratio, (B-V) rest frame 
color, Sersic index, radius, LW age, 
MW age, 
local galaxy density, and stellar mass.\footnote{Here the galaxy local density has
been computed from the circular area that, in projection on the sky,
encloses the 5 nearest projected neighbours brighter than $M_V \leq
-19.85$ within $\pm1000 \, \rm km \, s^{-1}$, minimizing survey edge
effects using the SDSS and 2dFGRS in the PM2GC surrounding regions (see
Vulcani et al. 2012 for details).} For comparison,
overplotted are also the distributions for non-SDG galaxies. 
Table~2 summarizes the mean values for SDGs.

Most of the SDGs lie on the color-magnitude red sequence.
%, as shown in Fig.~\ref{fig:allprops}. 
Their color distribution is similar to
that of the whole mass-limited sample, showing that being red is a
characteristic of galaxies in this mass range regardless whether they
are compact or not.

Our SDGs show the tendency to be flattened on average, with a 
$<b/a> = 0.48$, as might be expected given the high number of S0s.\footnote{The 
average b/a of high-z galaxies, found using Daddi et al. (2005), van Dokkum
et al. (2008) and Damjanov et al. (2009), is 0.63$\pm$0.23.}
Similarly, most of our SDGs have Sersic indices lower than 4, with
$<n>=2.8$ and a distribution intermediate between disky and spheroidal
galaxies, as also high-z compact quiescent galaxies do (Wuyts et al. 2010,
van der Wel et al. 2011).

Most SDG radii are between 1 and 2 kpc, with a mean $<R_e>=1.4$.
Stellar masses can be as high as $10^{11.6}$, but most of them are
below $10^{11}$ with a mean $10^{10.8}$ ($6 \times 10^{10} M_{\odot}$).
SDGs are intrinsically rather bright, with a mean absolute V magnitude
of -20.9.

\begin{table}
\begin{center}
\begin{tabular}{ccccc}
%\hline
   &  \multicolumn{2}{c}{PM2GC}  & \multicolumn{2}{c}{WINGS B-band} \\
Quantity & Value & RMS error & Value & RMS error \\
\hline
\hline
\# SDGs     			& 44 	& 7           & 51 	& 7   \\
$\langle\re\rangle$       	& 1.45  & 0.26        & 1.57  & 0.34   \\
$\langle n\rangle$ 		& 2.8 	& 0.6         & 3.1 	& 0.8   \\
$\langle b/a\rangle$ 		& 0.48 	& 0.13        & 0.65 	& 0.16    \\
$\langle\sm\rangle$ $M_{\odot}$		& $6.0\per10^{10}$ & $1.9\per10^{10}$ & $9.1\per10^{10}$ & $3.6\per10^{10}$ \\
$\langle V_{abs}\rangle$ 	& -20.87& 0.42        & -20.68& 0.38   \\
$\langle\rm{Lw-age}\rangle$ 	& 5.45 	& 1.87        & 9.64 	& 2.10   \\
$\langle\rm{Mw-age}\rangle$ 	& 9.25 & 1.08         & 11.95 & 1.39   \\
Ellipticals frac.  		& 22.7\% 	& 7.2\%  & 29.1\% 	& 7.8\%  \\
S0s frac.          		& 70.5\% 	& 12.7\% & 62.0\% 	& 10.7\%  \\
Late-type frac.  		& 6.8\% 	& 3.9\%  & 8.8\% 	& 4.4\%  \\
\hline
\hline 
\end{tabular}
\caption{Characteristic numbers of PM2GC SDGs, compared to WINGS. 
Errors are derived from
Poissonian   statistics for     counts and fractions,   
and are     RMS    for  other
quantities.\label{tab:num}}
\end{center}
\end{table}

\begin{figure}
\centering
\includegraphics[scale=0.4]{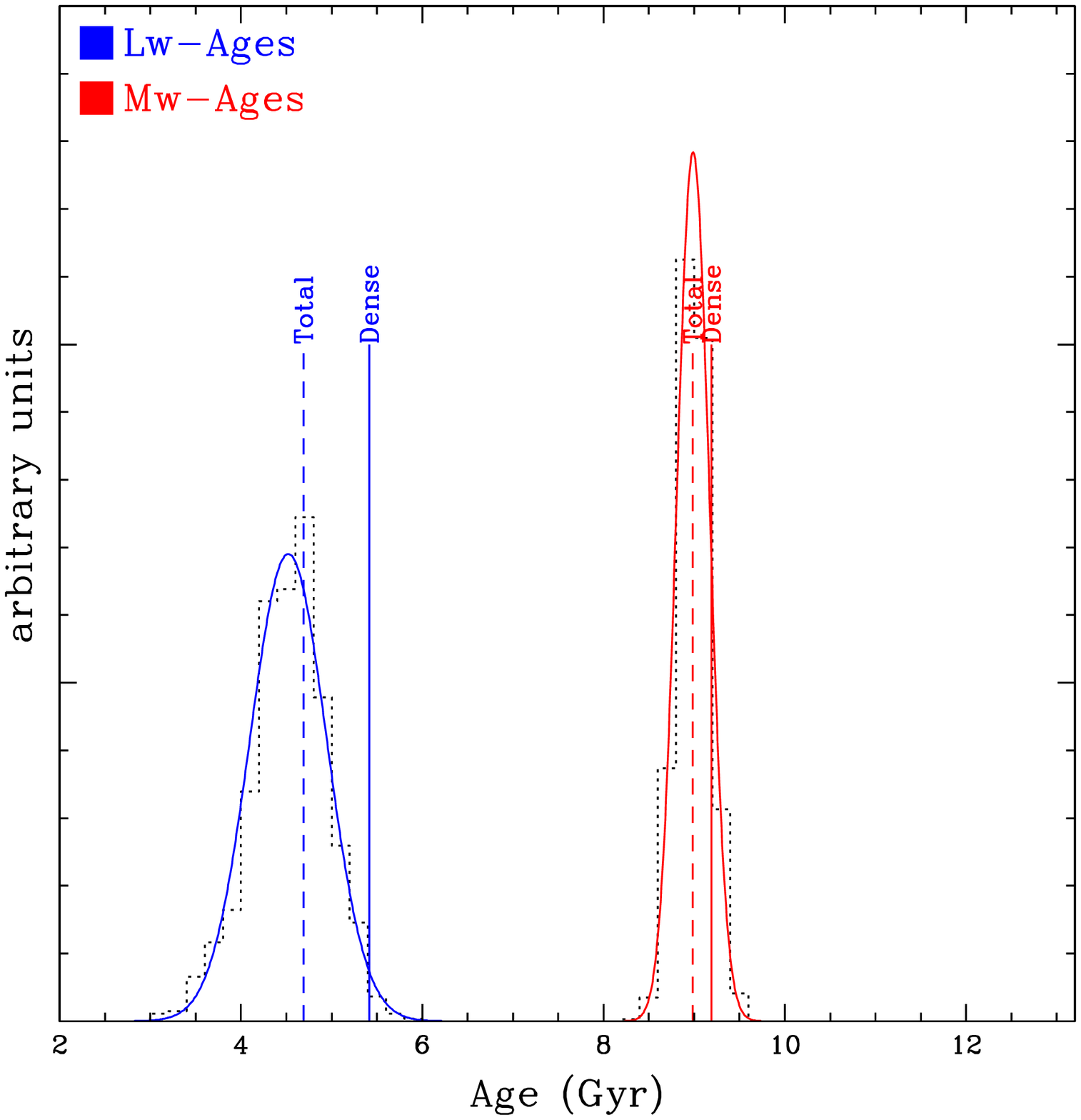}
\caption{Distributions of median luminosity-weighted
ages  (left    blue histogram)  and   mass-weighted   ages (right  red
histogram)  of 1000  random  extractions  of non-superdense early-type
galaxies  {\it with the same  mass distribution of SDGs}. The vertical
lines are the median ages of SDGs (solid  lines) and of non-superdense
galaxies (dashed lines) with their intrinsic mass distribution.
\label{fig:mcarlo}}
\end{figure}

PM2GC SDGs have an intermediate median LW age (5.4$\pm1.5$Gyr) and a old
median MW age (9.2$\pm0.8$ Gyr), showing that the bulk of the mass has
an age typically 4 Gyr older than the luminosity-weighted age.  The
great majority (93\%) of SDGs galaxies have
LW age $> 1.5$ Gyr. We have also inspected their
spectra to search for the presence of emission lines, and found that
only about 10\% of the SDGs have a significant emission ($>2-3$ \AA)
in any of the strongest lines ([OII] or $\rm H\alpha)$, never stronger
than 15\AA $\,$ in [OII].

The issue of the variation of galaxy sizes with
their stellar population ages 
is particularly relevant when investigating the size
evolution of galaxies. In addition to the well-known
increase of mean stellar age with galaxy mass, some works, and in particular
the cluster low-z 
work of V10, have found that, at fixed mass,
smaller (more compact) galaxies are older. 
This is confirmed by several works both at high- and low-z
(Saracco et al. 2009, Shankar \& Bernardi 2009, 
Kriek et al. 2009, Williams et al. 2010, Cappellari et al. 2012b),
although for example Trujillo et al. (2011) find no such a trend.
% in field galaxies at low- and high-z. 
Since most of the high-z studies select
massive galaxies that are {\it passive}, whose stellar populations
are already quite old at high-z, they would tend to select
the smallest, i.e. more compact galaxies, with a strong selection bias
(V10, Saracco et al. 2011).

To quantify to what extent, on average, PM2GC SDGs are older than non-compact
galaxies of similar masses, we used the Monte carlo technique to extract 1000
random samples of ``normal'' galaxies with the same mass distribution
of the SDGs. In Fig.~\ref{fig:mcarlo}, we plot the distributions of the median
LW and MW ages of these 1000 samples. The Monte Carlo simulation shows
that compact galaxies on average are $\sim 0.7$ and $\sim 0.2$ Gyr older 
in LW
and MW age, respectively, than normal galaxies of the same mass (the median
LW and MW ages for normal galaxies are 4.7$\pm 1.8$ and 9.0$\pm 0.8$ Gyr).
Again, this is suggesting that in some way age is related to compactness, 
in addition than mass, although the effect is modest for the LW-age
and barely appreciable for the MW age. 

We will come back to this point
and compare with LW and MW ages of cluster SDGs in \S4, discussing
the relevance of this effect for evolutionary studies.

\subsection{Velocity dispersions and dynamical masses}

\begin{figure}
\centering
\includegraphics[scale=0.4]{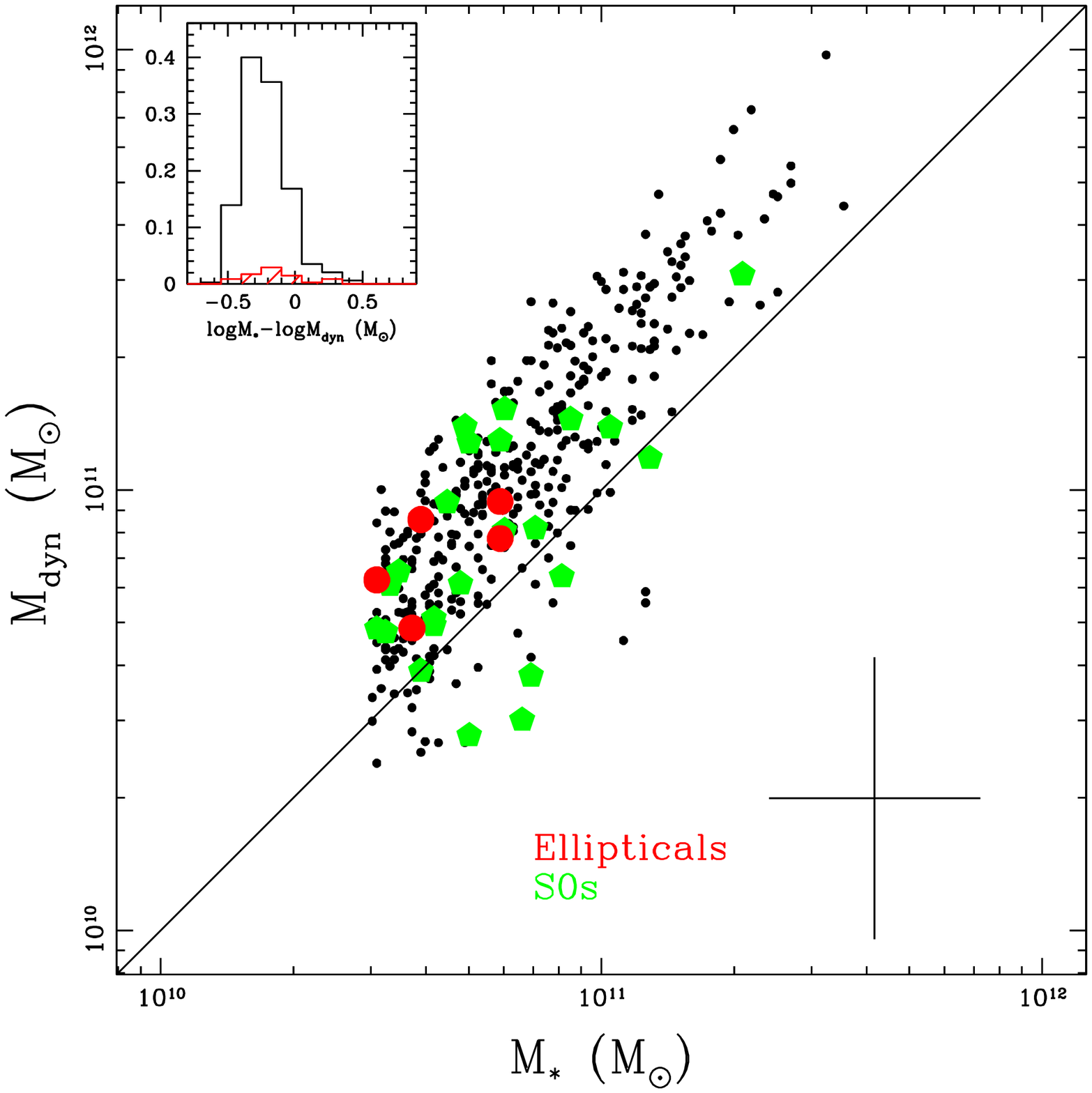}
\caption{The dynamical versus stellar mass relation for all PM2GC early-type (Es+S0s)
galaxies (small black dots) and for SDGs (large coloured circles - red for ellipticals and green for S0s).
The typical errorbar is shown in the bottom right corner.
The distribution of logarithmic difference between the two masses is shown in the inset
for all galaxies and for SDGs (empty black and shaded red histograms, respectively).
\label{fig:mdyn}}
\end{figure}

As a further check of our SDGs mass and density estimates, 
we use the galaxy central velocity dispersions available for
a subset of 570 galaxies in our mass-limited sample from the Sloan
DR7 database.
We compute the dynamical mass of all early-type (ellipticals + S0s)
galaxies in our mass-limited sample following 
%Cappellari et al. (2006, 2012):
the formulation by Bertin et al. (2002):

\begin{equation}
M_{dyn} = (K(n)\sigma^2 R_e)/G 
%M_{dyn} = (5\sigma^2 R_e)/G 
\end{equation}

where%

\begin{equation}
K(n) = [73.32/(10.465+(n-0.95)^2)] + 0.954
\end{equation}

using the GASPHOT Sersic index $n$ and circularized radius $R_e$
and the DR7 velocity dispersion
$\sigma$ corrected to $R_e$ as suggested in 
Cappellari et al. (2006). Using a fixed K(n)=5 as recommended
by Cappellari et al. (2006,2012) does not change the results.
%Jorgensen et al. (2005).

The comparison of virial and stellar mass estimates is shown in
Fig.~\ref{fig:mdyn}.  For the great majority of
galaxies, as well as the great majority of SDGs, the dynamical mass is
higher than the stellar mass by a factor that is very similar to values
found in the literature (Taylor et al. 2010b).  
The median difference $log M_{\star} - log M_{dyn}$ is -0.24dex.
There are very few
cases with stellar mass estimates higher than dynamical estimates,
compatible within the errors.  
% for all early-type galaxies and -0.13 for SDGs.  
%The smaller offset for SDGs is probably due to the
%high fraction of S0s (rotationally supported galaxies) among SDGs, for
%which the dynamical mass estimate method may be inappropriate.
The comparison between dynamical and stellar masses
confirms the validity of the mass estimates we have used so far
and confirms there is no problematic issue regarding our mass estimates.

\begin{figure}
\centering
\includegraphics[scale=0.4]{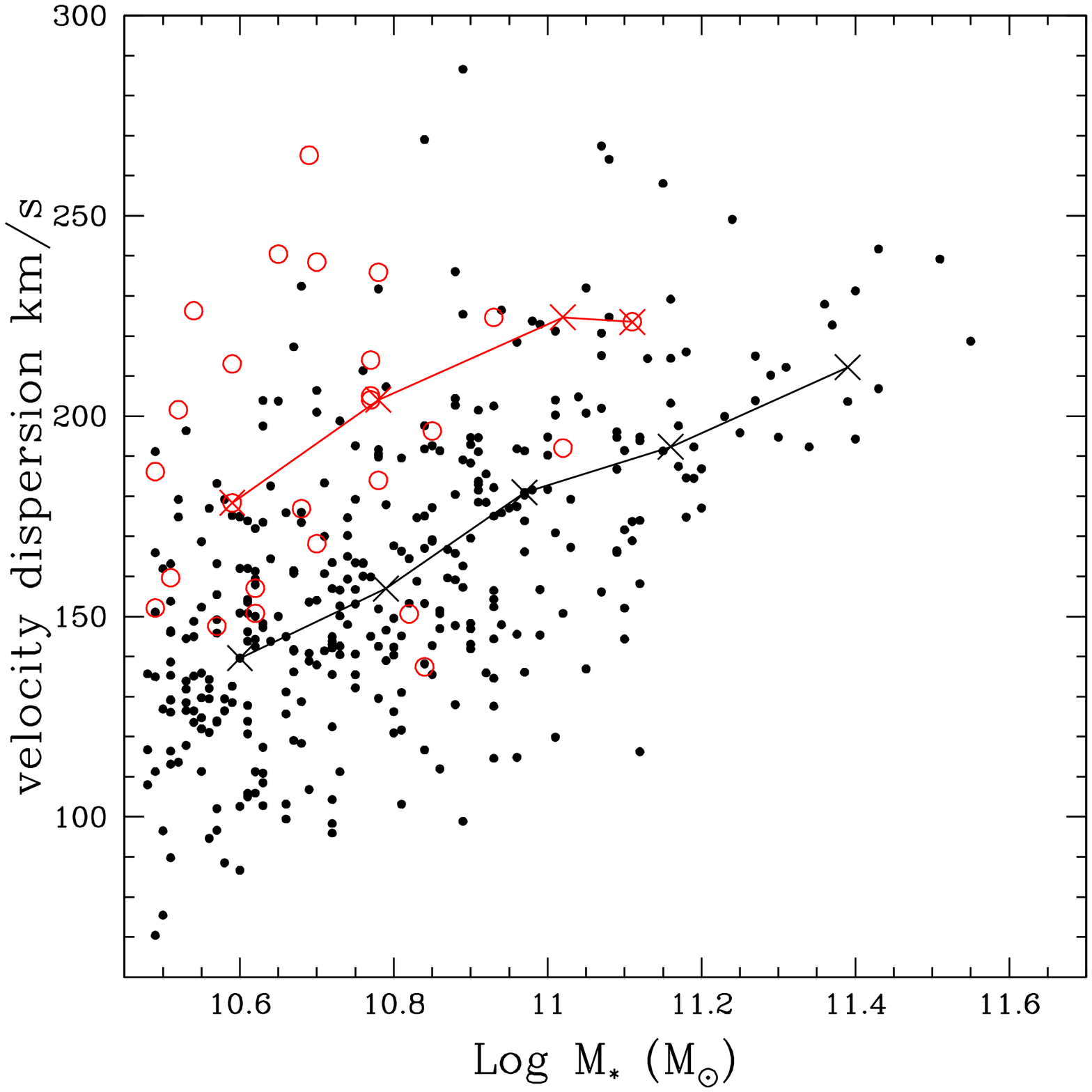}
\caption{Velocity dispersion versus stellar mass relation for non-SDG 
early-type galaxies (ellipticals + S0s, black) 
and superdense early-type galaxies (red) for which a velocity dispersion 
measurement is available from Sloan. The black line is the median relation
for non-SDGs, red for SDGs only. Velocity dispersions have all been corrected
to $R_e$ as for Fig.~\ref{fig:mdyn}.
\label{fig:sigma}}
\end{figure}

As a further, final test we check whether the velocity dispersions of SDGs
are higher than non-SDG galaxies of similar mass, as expected if they
are indeed more compact than the normal population.
Figure~\ref{fig:sigma}
shows that our SDGs (red symbols) have significantly higher velocity
dispersions than the average non-compact galaxies of similar stellar
mass. The
median relations for SDGs and non-SDG galaxies are also shown.
For SDG galaxies, the median offset of $log R_e$ from the mass-radius 
relation is
-0.4dex, while the median $\Delta log \sigma$ is 0.1dex.
These values
are very similar to the corresponding ones found by Taylor et
al. (2010) for their SDSS sample (-0.38 and 0.12dex, respectively).
The dynamical mass estimates and the velocity dispersions of our SDGs
are therefore consistent with the fact that these are massive and 
compact objects.

\section{Environment}

As explained in \S2, within the PM2GC survey, Calvi et al. (2011) have 
identified different environments: single galaxies, binary systems, and 
groups with a wide range of velocity dispersions. These environments
span a range of dark matter halo masses, of the order of 
$10^{12.5}$ to $10^{15} M_{\odot}$. We can therefore quantify the occurrence 
of SDGs in different environments. 

We find that the SDG fraction is always around 4\%$\pm$1\%,
hence indistinguishable, in groups, binaries and single
galaxies. However, the fraction rises to 14$\pm7$\% in PM2GC systems
with a velocity dispersion $>500 \rm \, km \, s^{-1}$. 
Splitting groups further into $<300 \rm \, km \, s^{-1}$
and $ 300-500 \rm \, km \, s^{-1}$ we do not detect 
a difference in SDG fraction, but uncertainties are large
and there might be a more gradual transition in SDG fraction
with environment which we do not detect due to small number statistics.

In constrast with high-z results finding a trend of sizes with local
galaxy overdensity (Cooper et al. 2012), there is instead no
preference for SDGs to be in high density regions within the PM2GC, as
shown in Fig.~\ref{fig:allprops}.

\subsection{Field versus cluster superdense galaxies}

SDG galaxies in nearby clusters were studied by the WINGS survey 
(V10). WINGS is a photometric and spectroscopic
multiwavelength survey of 77 X-ray selected galaxy clusters at $z=0.04-0.07$
with a cluster velocity dispersion typically in the range 500-1200 $\rm km \,
s^{-1}$ and $L_X= 0.2-5 \times 10^{44} \rm \, erg \, s^{-1} $
(Fasano et al. 2006, Cava et al. 2009).

Effective radii and galaxy morphologies
were obtained from V-band images with the same tools used here,
GASPHOT and MORPHOT, respectively (Pignatelli \& Fasano 2006, 
Fasano et al. 2012). Stellar galaxy masses were estimated
both using the method adopted in this paper and with the
spectrophotometric model used to derive stellar population ages, finding 
excellent agreement between the two methods (Fritz et al. 2011, Vulcani et al. 2011).
LW and MW ages were derived from optical spectra with the same method
and spectrophometric model we used for the PM2GC (Fritz et al. 2007, 2011).

The WINGS and PM2GC surveys, together, explore the whole range of
environmental conditions, from massive clusters to groups, binary and
isolated galaxies.  All the tools and methods
of analysis are the same for WINGS and the PM2GC, allowing a
homogeneous comparison between massive clusters and general field in
the nearby Universe.

For the purposes of this paper, the only relevant difference between
the two surveys is that the size measurements in V10 were performed on V-band WINGS images, instead of
B-band as in the PM2GC.\footnote{The consistency between V and B-band
WINGS morphological classifications is discussed in Calvi et
al. 2012.}  To obviate this problem, we have employed GASPHOT
to obtain effective radii from WINGS B-band imaging
(Bindoni et al. in prep.), which was taken 
with the same instrument (WFCAM/INT) and similar depth of the MGC
images we used in this paper.

\begin{figure}
\centering
\includegraphics[scale=0.55]{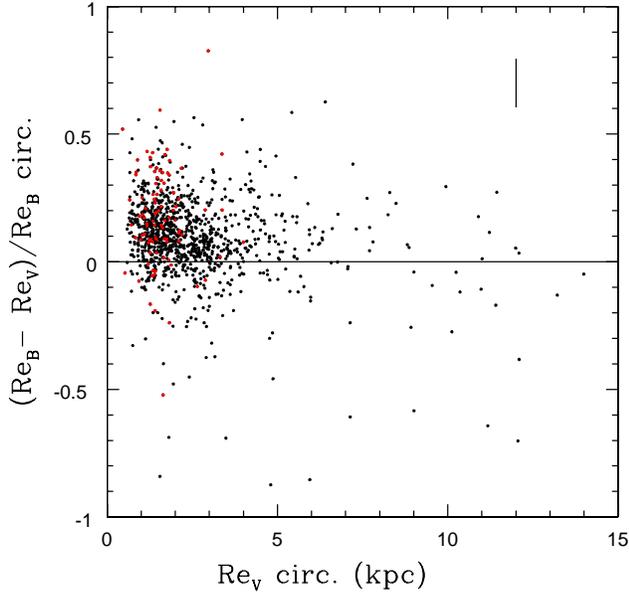}
\caption{Differences in circularized $R_e$ estimates in the B and V bands for WINGS galaxies.
Red circles are V-band SDGs. The errorbar in the right upper corner represents
the median relative error in the B-band $\Delta R_e/R_e = 0.1$.
\label{fig:bwings}}
\end{figure}

It turned out that the difference between the two filters is small, as shown
in Fig.~\ref{fig:bwings}, but should not be neglected.
The median difference between the WINGS B-band $R_e$ and the V-band $R_e$
is $0.20\pm0.19$ kpc for all galaxies, equivalent to a median
$\Delta R_e/R_e \sim 0.1$, and $0.29\pm0.30$ kpc 
for SDGs with median $\Delta R_e/R_e \sim 0.17$. GASPHOT was also run on WINGS 
K-band images (Valentinuzzi et al. 2009,
Bindoni et al. in prep.) and  we find a small but clear trend of galaxy sizes
with wavelength as radii increase going from K to V to B.
In the following, we only use B-band WINGS results.

This small, but systematic offset has an effect on the SDG fraction.
Adopting the usual criteria to define SDGs (equations 3 and 4), B-band
SDGs in WINGS are 11.8$\pm$1.7\% of all galaxies more massive than our adopted
mass limit. This fraction is about half of that found in the V band
for the same limit (22$\pm2$\%, V10).  The B-band SDGs 
are essentially a subsample contained in the V-band SDG sample.

The WINGS SDG proportion is consistent with the rough estimate obtained
for $>500 \rm \, km \, s^{-1}$ systems from the PM2GC ($14\pm7$\%,
\S3.1).  SDGs are therefore {\it proportionally} much more numerous in
clusters than in other environments, representing a significant
fraction of the cluster massive galaxy population but a factor of 3 smaller
fraction in the field.

Let us now compare the number density of SDGs in the field and in
clusters, summarized in Table~3.  As said in \S3.0, the number density
of SDGs in the PM2GC is $4.3 \times 10^{-4} \, \rm h^3 \, Mpc^{-3}$.
Considering only the {\it oldest} SDGs, those that have been quiescent
at least since z=1.0 (=have stopped forming stars at least 1 Gyr before
z=1, corresponding $\sim$ to a LW age $> 9$ Gyr), the number
density in the PM2GC is almost an order of magnitude smaller, $5.9 \times
10^{-5} \rm h^3 \, Mpc^{-3}$.\footnote{A direct comparison with published 
values of the number density
of high-z compact galaxies is unfortunately unfeasible, because
compactness criteria do not match and/or the largest high-z samples
are affected by the $z=1.6$ overdensity in the GOODS-S field
(Cassata et al. 2011, Barro et al. 2012).
A careful comparison of the low-z and high-z number densities 
will be very valuable to assess what fraction of compact galaxies
have survived till today (Renzini 2012).}

The WINGS B-band SDGs account for a number density of {\it at least}
$1.7 \times 10^{-5} \rm h^3 \, Mpc^{-3}$ at $z=0.04-0.07$.  This
number has been computed by multiplying the average SDG number per
cluster among the 21 clusters considered in this study, corrected for
spectroscopic incompleteness, by the total number of clusters in the
WINGS survey, and dividing by the whole volume of the Universe between
$z=0.04$ and $z=0.07$. Considering the whole volume and not just the cluster
volume it is like assuming there are no SDGs outside of clusters,
obtaining a hard lower limit on the SDG number density at these redshifts.
We have not taken into account the fact that
WINGS images are sampling the cluster populations as far as $\sim 0.6 \times 
R_{200}$
 for all clusters, while they reach the
virial radius for a small subset of clusters.  Hence, the numbers
above include only galaxies in a fraction of the cluster volume, and
the WINGS number density estimates are hard lower limits to the total
SDG density accounted for by clusters.

WINGS SDGs are generally very old, with a median LW age of 9.5Gyr,
and the number density of LW age$>9$ Gyr WINGS SDGs is {\it at least}
(for the reasons given above)
$1.0 \times 10^{-5} \rm h^3 \, Mpc^{-3}$. 

Although we cannot obtain a precise estimate of the fraction of
low-z SDGs that reside in clusters, due to the reasons outlined
above, 
we can conclude that, among {\it old} SDGs in the local Universe, a
significant fraction ($>17$\% = $1.0 \times 10^{-5}/5.9 \times 10^{-5}$) 
must be found in clusters today.

\begin{table}
\begin{center}
\begin{tabular}{lc}
%\hline
 Criteria & Number density \\ %& Literature \\
 & $\rm{h^3 \, Mpc}^{-3}$ \\ %& $\rm{h^3 \, Mpc}^{-3}$ \\
\hline
\hline
General field SDGs (PM2GC)       			& $4.3\pm0.06 \, \, 10^{-4}$ \\
PM2GC SDGs quiescent today (LWage $>1.5$Gyr)            & $4.0\pm0.06 \, \, 10^{-4}$ \\
PM2GC SDGs quiescent z=1.0 (LWage $>9$Gyr)              & $5.9\pm0.6 \, \, 10^{-5}$ \\ 
WINGS ``cluster'' B-band SDGs                           & $>1.7\pm0.09 \, \, 10^{-5}$ \\
WINGS B-SDGs quiescent z=1.0 (LWage$>9$Gyr)             & $>1.0\pm0.06 \, \, 10^{-5}$ \\
\hline
\hline 
\end{tabular}
\caption{Number densities of SDGs and quiescent galaxies. 
%Literature data:         Bez=\citet{bezanson09},
%Cimatti=\citet{cimatti08},Wuyts=\citet{wuyts09}.  
Errors are derived from Poissonian statistics.
\label{tab:nden}}
\end{center}
\end{table}

\subsection{Cluster SDG properties}

Table~2 presents the average properties of WINGS B-band
SDGs.\footnote{We note that these properties 
are similar to those of V-band SDGs
given in Table~3 of V10. The distributions of
SDG quantities are also similar to Fig.~6 in V10, and the latter can be compared with Fig.~6 in this paper.
While the SDG fraction depends on the band used for their selection, the SDG properties don't.}
They have been computed weighting the values for spectroscopic
incompleteness, as in V10.

Their mean effective radius, Sersic index, axial ratio and absolute V
luminosity are rather similar to the mean values of PM2GC SDGs. As in
the field, the majority of cluster SDGs are S0s (62\%) or ellipticals
(29\%).  There is a tendency for cluster SDGs to be on average more
massive than field SDGs (9 versus 6 $\times 10^{10} M_{\odot}$), but
the most striking difference is in the stellar population ages.

Cluster SDGs are on average 4 Gyr older than field SDGs in LW age, and
about 3 Gyr older in MW  age. The age distributions for B-band WINGS
SDGs are not dissimilar from those given in Fig.~6 of V10 for V-band selected SDGs and can be compared with Fig.~6 in
this paper. Only 3 out of 51 cluster SDGs have a LW age younger than 6
Gyr, while about half of the field SDGs are this young.

The relation between galaxy mass, size and stellar age as a function
of environment will be presented in detail in the next section, where
we draw our conclusions regarding the evolution of high-z
massive compact galaxies and their descendants in the local Universe.

\begin{table}
\begin{center}
\begin{tabular}{lcc}
%\hline
%\hline
\hline
 $\log_{10}(\sm/\msol)$ & \multicolumn{2}{c}{$\log_{10}(\re/\rm{kpc})$} \\
\hline
 & Late Types & \\
% & \multicolumn{2}{c}{Late Type galaxies} \\
 &  $T>0$ & $ n < 2.5$ \\ 
\hline
$10.00_{-0.07}^{+0.05}$  &  $0.34_{-0.11}^{+0.11}$  & $0.23_{-0.11}^{+0.11}$  \\
$10.27_{-0.08}^{+0.03}$  &  $0.42_{-0.13}^{+0.08}$  & $0.31_{-0.17}^{+0.12}$ \\
$10.53_{-0.07}^{+0.06}$  &  $0.48_{-0.15}^{+0.08}$  & $0.39_{-0.16}^{+0.11}$ \\
$10.88_{-0.08}^{+0.07}$  &  $0.65_{-0.20}^{+0.13}$  & $0.46_{-0.16}^{+0.15}$ \\
$11.10_{-0.05}^{+0.08}$  &  $0.84_{-0.19}^{+0.10}$  & $0.58_{-0.08}^{+0.13}$ \\
\hline
 & Early Types & \\
% & \multicolumn{2}{c}{Early Type galaxies}\\
 &  $\rm T \leq 0$ & $n \geq 2.5$ \\  
\hline
$9.92_{-0.05}^{+0.04}$   &  $0.19_{-0.11}^{+0.10}$  &  $0.19_{-0.13}^{+0.17}$ \\
$10.11_{-0.06}^{+0.05}$  &  $0.19_{-0.09}^{+0.14}$  &  $0.24_{-0.09}^{+0.14}$ \\
$10.30_{-0.06}^{+0.06}$  &  $0.30_{-0.14}^{+0.10}$  &  $0.31_{-0.13}^{+0.13}$ \\
$10.49_{-0.04}^{+0.05}$  &  $0.35_{-0.14}^{+0.13}$  &  $0.36_{-0.11}^{+0.13}$ \\
$10.70_{-0.05}^{+0.05}$  &  $0.39_{-0.10}^{+0.15}$  &  $0.42_{-0.12}^{+0.14}$ \\
$10.90_{-0.05}^{+0.06}$  &  $0.49_{-0.11}^{+0.12}$  &  $0.51_{-0.12}^{+0.17}$ \\
$11.06_{-0.02}^{+0.06}$  &  $0.69_{-0.15}^{+0.14}$  &  $0.69_{-0.15}^{+0.17}$ \\
$11.29_{-0.04}^{+0.06}$  &  $0.65_{-0.13}^{+0.17}$  &  $0.72_{-0.16}^{+0.13}$ \\
$11.48_{-0.06}^{+0.05}$  &  $0.80_{-0.09}^{+0.14}$  &  $0.80_{-0.09}^{+0.14}$ \\
$11.65_{-0.03}^{+0.05}$  &  $1.03_{-0.35}^{+0.05}$  &  $0.92_{-0.14}^{+0.16}$ \\
%$11.56_{-0.00}^{+0.00}$  &  $0.89_{-0.00}^{+0.00}$  &  $0.89_{-0.00}^{+0.00}$ \\
\hline
 &  All galaxies & \\ 
\hline
$9.93_{-0.06}^{+0.03}$   &  $0.22_{-0.11}^{+0.10}$  &   \\
$10.10_{-0.05}^{+0.05}$  &  $0.23_{-0.10}^{+0.15}$  &   \\
$10.29_{-0.05}^{+0.06}$  &  $0.33_{-0.14}^{+0.11}$  &   \\
$10.49_{-0.04}^{+0.05}$  &  $0.37_{-0.13}^{+0.13}$  &   \\
$10.70_{-0.04}^{+0.05}$  &  $0.43_{-0.13}^{+0.11}$  &   \\
$10.90_{-0.06}^{+0.05}$  &  $0.51_{-0.12}^{+0.16}$  &   \\
$11.06_{-0.02}^{+0.07}$  &  $0.69_{-0.15}^{+0.17}$  &   \\
$11.29_{-0.04}^{+0.07}$  &  $0.69_{-0.14}^{+0.14}$  &   \\
$11.44_{-0.02}^{+0.09}$  &  $0.87_{-0.16}^{+0.15}$  &   \\
$11.68_{-0.04}^{+0.00}$  &  $0.92_{-0.14}^{+0.16}$  &   \\
%$11.56_{-0.00}^{+0.00}$  &  $0.89_{-0.00}^{+0.00}$  &   \\
\hline 
\end{tabular}
\caption{WINGS mass-radius relations in the B band. The values are logarithm of
the median estimates, errors are the lower  and upper quartiles of the
distributions.\label{tab:wlocal}}
\end{center}
\end{table}

\section{The mass-size relation: Descendants and progenitors}

In the previous section we have seen that the incidence of
SDGs and the ages of their stellar populations
are different in clusters and in the general
field. 
In this section we wish to address a more general question:
how does the relation between galaxy mass, size and stellar
age depend on environment, and what can this tell us
about the local Universe descendants of the high-z massive
and compact galaxies.

\begin{figure*}
\centering
\includegraphics[scale=0.6]{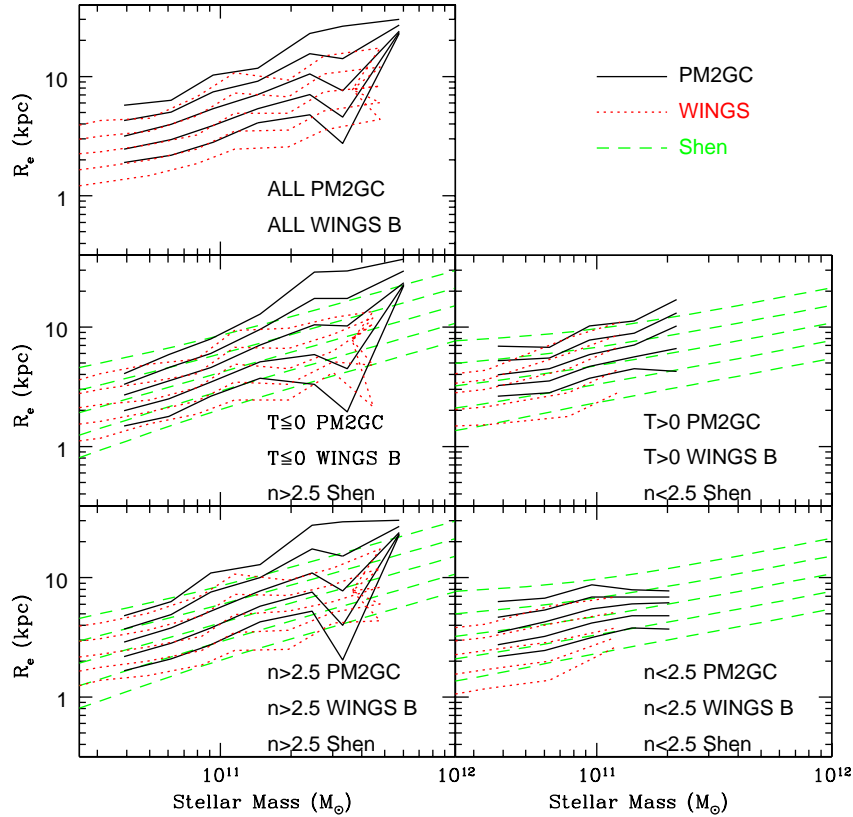}
\caption{Comparison of the median mass-size relation found in the PM2GC
(continuous black lines),
in WINGS (red dotted lines)
and those given by Shen et al. (2003) for the Sloan
$r$ band (green dashed lines). 
The comparison is done using: all galaxies (top left panel),
early-type ($T \leq 0$) PM2GC and WINGS galaxies with $n>2.5$ Shen's
galaxies (middle left), late-type ($T>0$) PM2GC and WINGS galaxies with 
$n<2.5$ Shen's galaxies (middle right), $n>2.5$ PM2GC, WINGS and
Shen's (bottom left), and $n<2.5$ PM2GC, WINGS and
Shen's (bottom right). The five lines for each dataset represent
the median (middle line), 1$\sigma$ and 2$\sigma$ relations.
All WINGS radii in this figure are based on B-band imaging. 
Shen et al. (2003) do not provide a mass-size relation for
all galaxies.
\label{fig:mreall}}
\end{figure*}

Let us first consider the mass-size relation.  Figure~\ref{fig:mreall}
compares the median mass-size relations obtained in the PM2GC and in
WINGS (B-band) for all galaxies and for galaxies of different types.
We consider separately early-type (T-type$\leq 0$)
versus late-type galaxies (T-type$> 0$), and galaxies with Sersic $n
\geq 2.5$ versus $n < 2.5$.  
The binned values are listed in Tables~1 and ~4. 
Figure~\ref{fig:mreall} includes also a comparison
with the relations given by Shen et al. (2003) ($n
\geq 2.5$ and $n < 2.5$) that are based on Sloan
$r$-band imaging. The following conclusions can be drawn from this figure:

a) There is an offset between the mass-size relation in the general
field and in clusters. Galaxies of a given mass in the field are
bigger than in clusters typically by about 1$\sigma$.  This effect is
visible for all galaxies, and also for each type of galaxies when
dividing them by morphological type or Sersic index, being negligible
only for late-type massive galaxies.  A KS-test can rule out the
hypothesis of similar size distribution for cluster and field galaxies
of a given mass with probabilities $2\times 10^{-8}$ at $10.5 \leq log
M_{\star} < 10.8$, $1.7\times 10^{-8}$ at $10.8 \leq log M_{\star} <
11.1$, $1\times 10^{-4}$ at $11.1 \leq log M_{\star} < 11.4$.
Considering only early-type galaxies, the probabilities are $6\times 10^{-3}$,
$9\times 10^{-5}$ and $3\times 10^{-5}$ in the three mass bins, 
respectively. We note that Huertas-Company et al. (2012b), instead, do not find
a change in the median mass-size relation of early-type galaxies
in the SDSS when comparing the general field and clusters.

This is all based on B-band imaging, therefore it cannot arise
from color gradient effects. We note that the WINGS B-band relation
is already shifted to larger radii compared to the V-band
relation for the same galaxies due to the wavelength dependence
of effective radius shown in Fig.~\ref{fig:bwings}.

b) The Shen et al. relation is shown here only to allow a comparison
with the several works that have used it.  When dividing by Sersic
index, the PM2GC relation for $n>2.5$ is shifted to larger radii
compared to Shen's by about 1$\sigma$, while the median relation is
rather consistent for $n<2.5$.  We remind that an offset to larger
radii is expected when using bluer bands. Moreover, the Shen et
al. work is known to be affected by systematic errors due to
background oversubtraction (Guo et al. 2009), a bug in the NYUVAC size
estimates used at the time that has been fixed in later NYUVAC
releases (Blanton et al. 2005) and by the SDSS incompleteness at low
redshifts (Taylor et al. 2010).

c) Dividing galaxies by morphological types and by Sersic index is of
course not fully equivalent. Differences can be observed
in the mass-size relations derived for late-type
galaxies (Tables~1 and ~4). This is to be expected, given that
dividing galaxies by Sersic index corresponds to include in the
$n<2.5$ group some early-type galaxies, that are generally smaller
than late-types and have the effect to decrease the average size value at a
given mass.

\begin{figure*}
\centering
\includegraphics[scale=0.8]{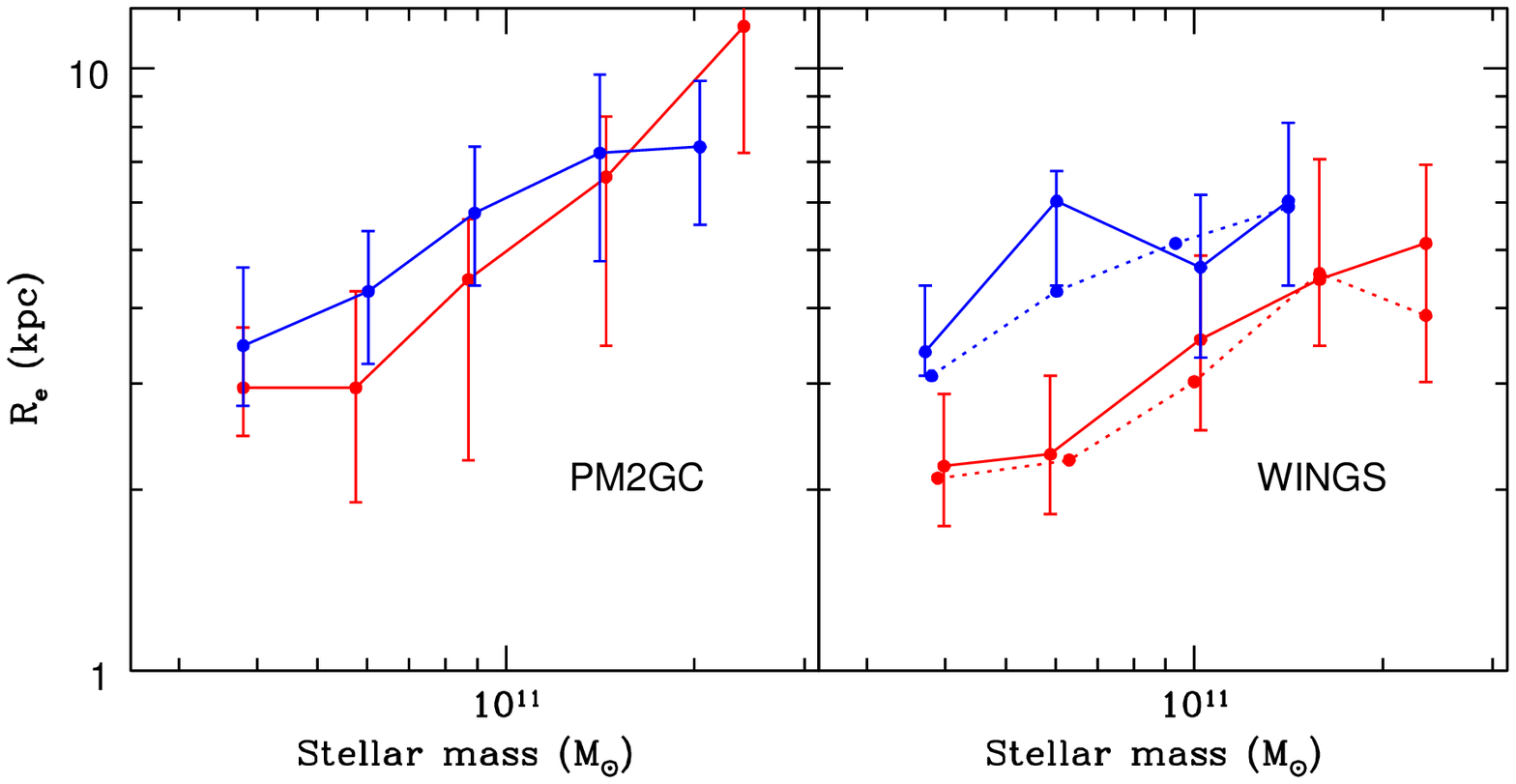}
\caption{Median mass-size relation for PM2GC (left) and WINGS (right)
galaxies with LW ages $>8$ Gyr (red continuous lines) and LW ages
$<4$ Gyr (blue continuous lines). Upper and lower quartiles
are shown as errorbars.
For WINGS, dotted lines refer
to LW ages $<6$ Gys (blue) and $>10$ Gyr (red). 
\label{fig:mreage}}
\end{figure*}

Let us now consider the effects of the stellar population age on the
mass-size relation.  In \S3.2 we have seen that the median LW age of
SDGs is 0.7 Gyr older than the LW age of a sample of non-compact
galaxies with the same mass distribution of SDGs. 
%The age gradient with size at
%any fixed mass in Fig.~7 is visible, but not outstanding.  
In constrast, V10 found cluster SDGs to be on
average 1.5 Gyr older than non-compact of the same mass.
%, and a much
%stronger dependence of stellar age on size than we find in the field.

Figure~\ref{fig:mreage} presents the median mass-size relation in the
PM2GC (left) and B-band WINGS (right) for galaxies of different LW
ages.  For the PM2GC we show galaxies younger than 4 Gyr (blue) and
older than 8 Gyr (red), that represent the youngest and oldest wings
of the age distribution. For WINGS, where the age distribution is
skewed to much older ages, we can plot galaxies younger than 4 or 6
Gyr (blue), and older than 8 or 10 Gyr (red).  

We conclude that both in the field
and in clusters, as one considers galaxies with older LW ages, the median
mass-size relation shifts to smaller sizes, but this effect is
more pronounced in clusters than in the field.
On average, in the mass range $3 \times 10^{10} - 2 \times 10^{11} M_{\odot}$,
the sizes of ``old'' (8+ Gyr in LW age) galaxies are a factor $\sim 1.2$
in the field and $\sim 1.5$ in clusters
smaller than those of ``young'' (4Gyr) galaxies.

This effect needs to be taken into account when comparing the sizes of
high-z and low-z galaxies to infer an evolution in the sizes of individual
galaxies. Our results show that selecting already passive galaxies at
high-z means selecting the most compact ones. Assuming they do not resume
the star formation activity at later times, the high-z sizes should be compared
with the sizes of the {\it oldest} (in terms of their
luminosity-weighted age) low-z galaxies, to avoid a significant progenitor
bias.

Moreover, we have shown that compact massive galaxies are much more
common in clusters than in the field today, and that cluster
SDGs are older than the field ones. It is therefore logical
to expect that a large number of the high-z massive galaxies selected
to be passive are found in clusters today. This is supported also by
the expectations obtained from hierarchical simulations, which we now
discuss.

\subsection{Theoretical expectations}

The median galaxy mass of the ensemble of high-z datasets used in this paper
is $M_{\star} = 10^{10.96} M_{\odot}$.
In simulations, the most massive galaxies at high redshifts are located in
the extreme peaks of the density distributions, and it is reasonable to
expect that many of them will end up in the most massive structures 
today. 

We use the Millennium Simulation (Springel et al. 2005), populated using the
semi-analytic model presented in De Lucia \& Blaizot (2007).
These simulations show that 40\% (28\%) 
of all galaxies more massive than
$M_{\star}=10^{11} M_{\odot}$ ($M_{\star}=5 \times 10^{10} M_{\odot}$)
at z=2-2.5 end up in haloes more massive
than $10^{14} M_{\odot}$ (WINGS-like clusters) by redshift 0.
This fraction decreases to 30\% (23\%) and 20\% (16\%) for galaxies with
$M_{\star} > 10^{11} M_{\odot}$ ($M_{\star}=5 \times 10^{10} M_{\odot}$)
at $z=1.6$ and $z=0.6$, respectively.

Interestingly, if one considers galaxies with 
$M_{\star}> 10^{11} M_{\odot}$ ($M_{\star}=5 \times 10^{10} M_{\odot}$)
{\it and} already passive 
($SFR< 1 M_{\odot}/yr$, all with a specific star formation rate
$SSFR = SFR/M_{\star} < 10^{-11} yr^{-1}$) 
at z=2.2, 1.6 and 0.6, the fraction
that will be in $\geq 10^{14} M_{\odot}$ clusters at z=0
rises to 59\% (45\%), 44\% (36\%) and 25\% (22\%), respectively.

Hence, a large fraction of high-z massive galaxies, and the majority
of high-z passive and massive galaxies have evolved
into today's cluster galaxies.
%studying high-redshift passive and massive
%galaxies means to a large extent 
%studying the progenitors of today's cluster galaxies, 
Clusters at low
redshift are the most natural location to look for the
descendants of the majority
of high-redshift massive and passive (and compact) galaxies.

From simulations it is also possible to obtain a prediction for the
fraction of high-z massive and passive galaxies that will end up being
the {\it central} galaxy of today's clusters, therefore the BCG of
clusters like WINGS's. 

Cluster BCGs are the only type of galaxies for which a large evolution
in size was required by our WINGS study (V10, see also Bernardi 2009).
The comparison of WINGS clusters and EDisCS clusters showed that the
mean size and mass of BCGs have respectively increased by factors of
$\sim 4$ and $\sim 2$ between $z \sim 0.7$ and $z \sim 0.04$
(Valentinuzzi et al. 2010b).

Simulations predict that 40\% (23\%), 26\% (14\%) and
7\% (3\%) of galaxies more massive than $log M_{\star}=10^{11}
M_{\odot}$ ($M_{\star}=5 \times 10^{10} M_{\odot}$) and passive ($SFR<
1 M_{\odot}/yr$) at z=2.2, 1.6 and 0.6 become BCGs of clusters at
z=0. Hence, a significant fraction of massive and passive (and compact) 
galaxies at $z \geq 2$ should be the progenitors of cluster
BCGs. As they are sitting in the best location of the Universe
for accreting other galaxies, their expected
minor merging rate is extremely high, and the time
between $z=2$ and $z=0$ could be sufficient for a colossal size growth 
to reach the $>20$ kpc sizes they have today.

\subsection{Progenitors and descendants:
the amount of size evolution in massive galaxies}

\begin{figure*}
\centering
\centerline{\hspace{2cm}\includegraphics[scale=0.58]{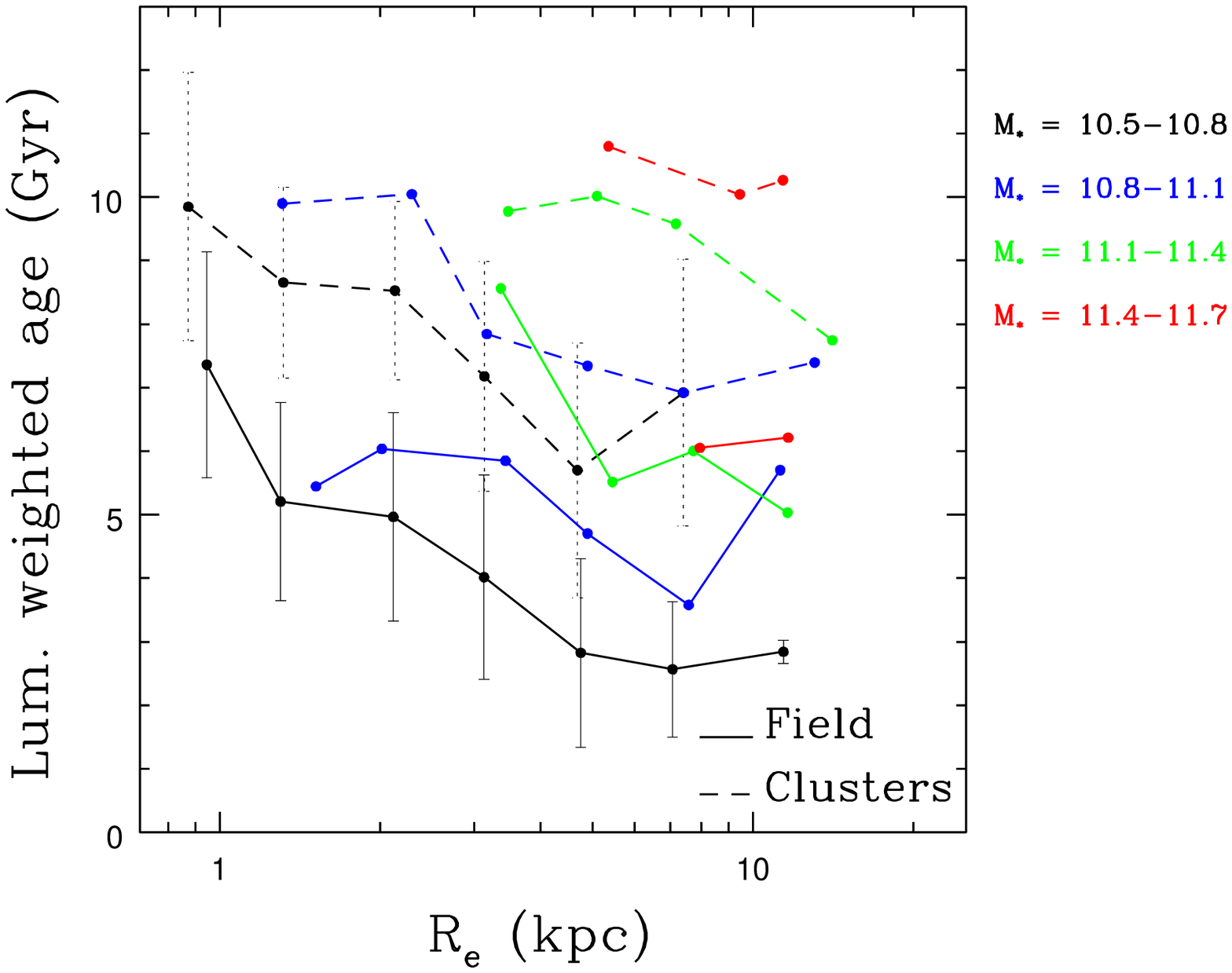}\hfill\hspace{-1cm}\includegraphics[scale=0.58]{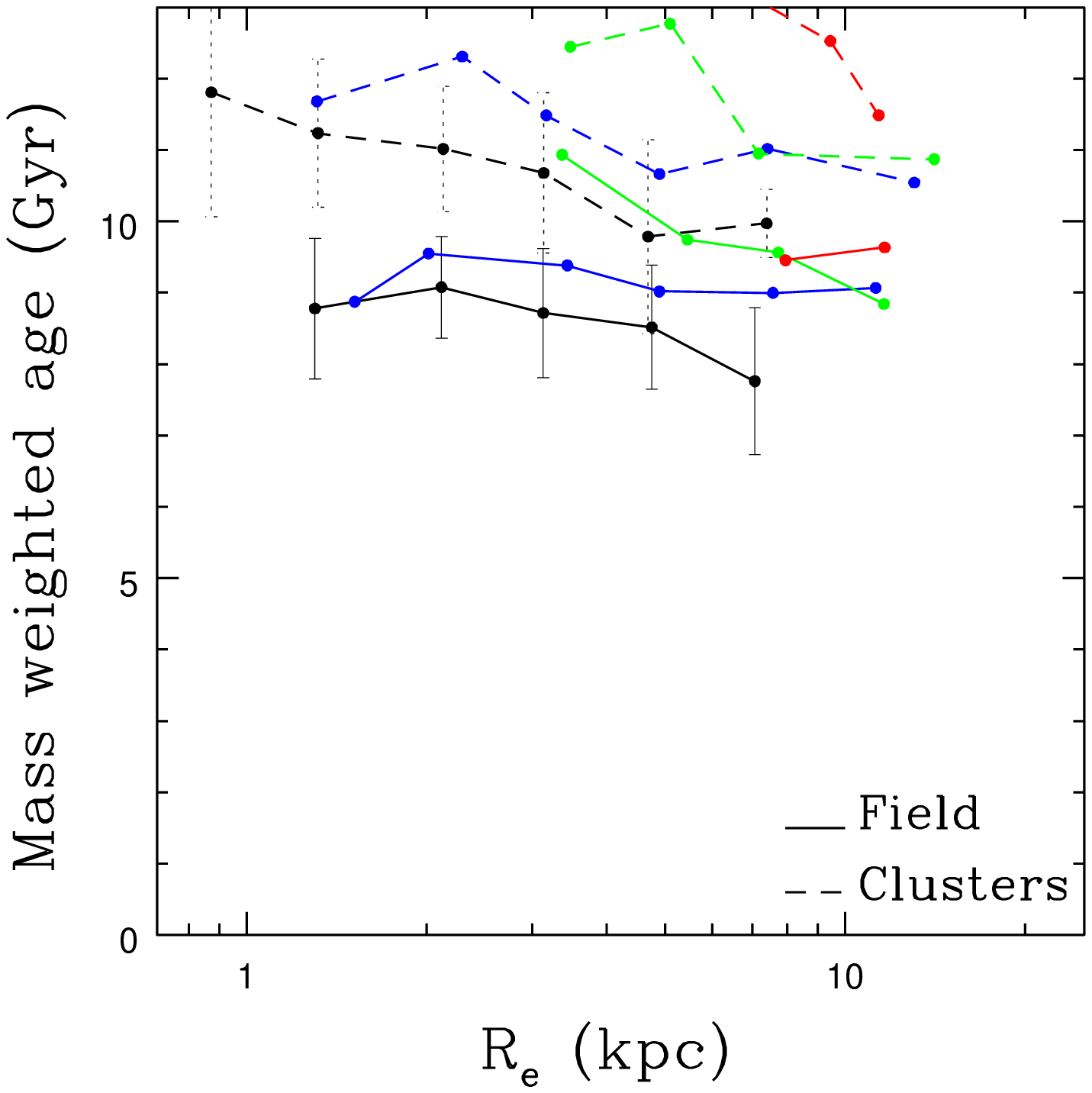}}
\caption{Median luminosity-weighted (left) and mass-weighted (right) age
of galaxies as a function of effective radius for four bins of galaxy masses.
Solid lines: field (PM2GC). Dashed lines: clusters (WINGS).
The rms of the median is shown as errorbars only for the lowest
mass bin, for clarity. Only points with more than 3 galaxies are
plotted.
\label{fig:summary}}
\end{figure*}

A summary of our results regarding
the relations between mass, size, age and environment
is presented in Fig.~\ref{fig:summary}. The most important trends
for the LW age are exemplified in the left panel:

a) at a given size, more massive galaxies have older LW ages, both in 
clusters and in the field. The mass of a galaxy has a strong influence on
the epoch when it stops forming stars.

b) at a given mass, galaxies with smaller radii have older LW ages, both in
clusters and in the field. The size (compactness) 
of a galaxy influences the epoch
when it stops forming stars.

c) at given mass and size, galaxies in clusters have older LW ages
than galaxies in the field. On top of the dependence on
mass and size, environment has a significant effect on
the epoch when galaxies stop forming stars.

When trying to quantify the impact of stellar-age
selection effects on the apparent evolution of the mass-size relation
of passive galaxies, the ``age'' to be considered is the
luminosity-weighted age, because the spectrophotometric properties
that are used by the high-z studies to select ``passive''
galaxies (color, absence of emission lines, SED age etc) are related
to the LW age, that reflects the epoch of the last star formation episode.
It is however interesting to analyse also the relations between mass, size, 
age and environment considering the MW age, related to the epoch
when the bulk of stars were formed. As shown in Fig.~\ref{fig:summary}:

d) at a given size, more massive galaxies have older MW ages, both in 
clusters and in the field, though the differences of age with mass
are smaller than for the LW age, at the level of 2 Gyr at most in our 
mass range. The mass of a galaxy has a significant influence on
the epoch when galaxies form the bulk of their stars.

e) the trends of MW age with radius at fixed mass are generally weaker
than for the LW age, especially for the two lowest mass bins.
The size of a galaxy seems to have a rather weak influence on when the bulk
of stars were formed. Our capability to detect such influence is
of course hampered here by the fact that the age resolution of
spectrophotometric models at such old ages is poor.

f) there is a noticeable dependence of MW age on environment, with
cluster galaxies of a given mass and size being older than their
field counterparts. On top of the dependence on galaxy mass and size,
environment plays a significant role in determining when galaxies
form the bulk of their stars.

The results we have presented highlight the necessity
to take into account the dependence of the mass-size relation 
on galaxy luminosity-weighted ages and environment
when studying the evolution of galaxy sizes. Failing to do so
has the effect of overestimating the evolution of the sizes of
individual galaxies.

\begin{figure*}
\centering
\includegraphics[scale=0.6]{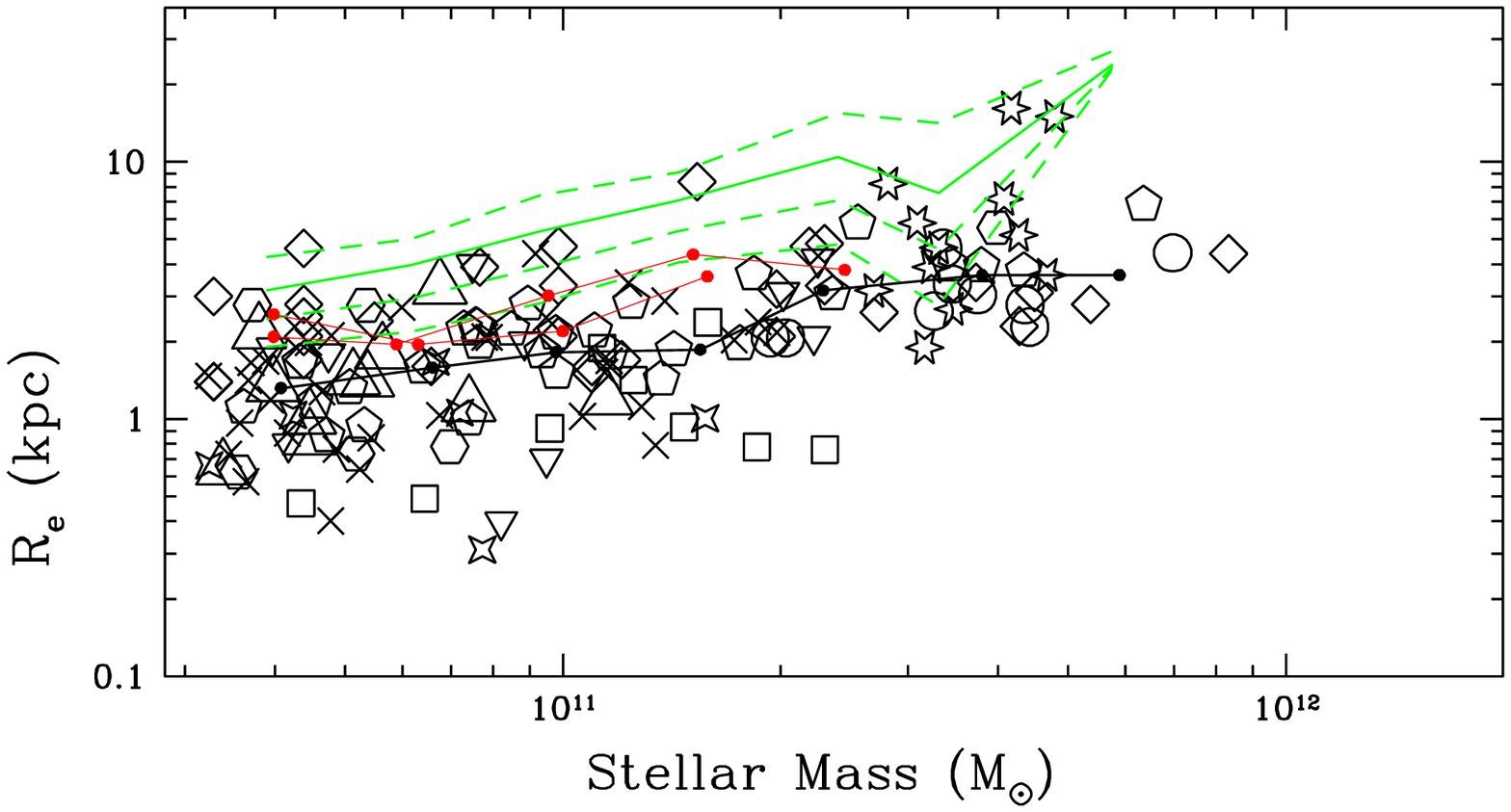}
\caption{Median mass-size relation at high-z and low-z.
Red lines: WINGS and PM2GC galaxies with LW ages
$\geq 10$ Gyr.
Black line: median to all high-z points.
Green lines: PM2GC all galaxies, with 1 and 2$\sigma$.
Large symbols indicate high-z datapoints.
\label{fig:allold}}
\end{figure*}

To quantify by how much the size evolution could be overestimated, 
we compare the high-z median mass-size relation
(solid black line in Fig.~\ref{fig:allold})
with that of {\it old} (LW age$ \geq 10$
Gyr) WINGS and PM2GC galaxies (red lines in figure). We find an average
evolution of 0.2-0.25dex, i.e. a factor 1.6-1.8, for WINGS and PM2GC
respectively.\footnote{We note that this amount of evolution is in
agreement with the factor 1.5 at most
 estimated by V10 comparing the sizes of galaxies
of different luminosity-weighted ages with galaxies at different
redshifts.} This factor represent by how much the median mass-size
relation shifts between high-z and low-z when considering the
most likely descendants of the high-z progenitors.
We do not attempt to assess the amount of evolution as a function
of galaxy mass (see e.g. Ryan et al. 2012 and Huertas-Company
et al. 2012), because this would require a mass-representative high-z sample.

Comparing instead the high-z relation with the PM2GC
relation for all galaxies with $n>2.5$ or all of today's passive galaxies
at face value, one would find an evolution of about 0.5dex, a
factor 3-3.2.  Therefore, environmental and selection effects can
account for half of the apparent effect.

We conclude that a significant fraction of the massive compact
galaxies found at high-z has evolved into today's cluster compact
galaxies.  When accounting for the age-selection effects and the
hierarchical assembly bias, the evolution of the high-z population in
size is smaller than would be inferred at face value considering as
descendants the whole passive low-z population regardless of age and
environment.  The size evolution of individual galaxies is modest, of
the order of a factor $1.6-1.8$ on average.

\section{Comparison with previous works}

Most of the high-z works use as local comparison the median mass-size
relation derived for $n>2.5$ galaxies by Shen et al. (2003) from Sloan
$r$-band imaging, finding several high-z points to lie at more than 2$\sigma$
from such relation.  In \S5 we have seen that our field relation for
$n>2.5$ galaxies is shifted by about 1$\sigma$ to larger radii
compared to the Shen et al. relation, and that such offset might
result from a combination of effects (bluer band used in this work,
systematic error in size estimates used by Shen, etc).

A few works have tried to identify massive compact galaxies, similar
to the high-z galaxies, at low redshifts.  Two works in particular
have searched for superdense galaxies in the nearby Universe, both of
them using Sloan-based size and mass measurements.

Trujillo et al. (2009) searched for galaxies with $M_{\star} > 8
 \times 10^{10}$ and $R_e \leq 1.5$ kpc in the Sloan at $0 < z < 0.2$
 using NYUVAC radii and mass measurements. After excluding stars, QSOs
 and cleaning the sample from those close to bright stars, in pairs
 and edge-on disks, they were left with 29 SDGs.  Applying Trujillo et
 al. (2009) criteria to define SDGs, we find 4 galaxies in our sample,
 out of 295 (1.4\%) more massive than Trujillo's limit.  As we find in
 this work and as V10 found in nearby clusters,
 Trujillo's SDGs are mostly fast rotators S0s (Ferre-Mateu et
 al. 2012, Trujillo et al. 2012).  The most intriguing result of their
 work is the fact that their SDG stellar population ages are young,
 suggesting that these galaxies have experienced recently enormous
 bursts of star formation (Trujillo et al. 2009, Ferre-Mateu et
 al. 2012).  In contrast, the stellar populations of our SDGs selected
 according to Trujillo's criteria are old: the MW age is always
 greater than 8 Gyr, the mean LW age is 6.3 Gyr and is $>5$ Gyr for all
 except one galaxy with LW age=1.2Gyr.

Another work searching for SDGs in the local Universe selected SDSS
red sequence galaxies with $M_{\star} > 10^{10.7} M_{\odot}$
with sizes less than half the size predicted by the Shen et al. (2003)
$n>2.5$ relation, excluding galaxies that were disk-like by visual inspection, 
showed marked asymmetries, were merging or affected
by bright stars (Taylor et al. 2010). We note that excluding visually
disk-like galaxies they probably would have excluded most of our sample.
Using de Vaucoulers DR7 radii and MPIA SDSS stellar masses, they found
63 galaxies at $0.066 < z < 0.12$. Using V10 criteria to define
SDGs, Taylor et al. get an SDG fraction of 1.3\% for red sequence galaxies.
If we use Taylor et al. (2010) criteria to define SDGs (red sequence
galaxies, $M_{\star}>10^{10.7} M_{\odot}$, $log(R_e)(kpc) < 0.56 * 
(logM_{\star} -9.84) -0.3)$, but without excluding disks,
we identify 13 galaxies out of 589 above this mass
limit, therefore a fraction equal to 2.2\%. Excluding S0s and spirals,
only 3 galaxies remain, a fraction 0.5\%.
In contrast with Trujillo's findings and in agreement with ours, Taylor's SDGs
have old stellar population, with LW ages typically in the range 6-10 Gyr.

Overall, our results agree with previous works in finding that
superdense galaxies are a small fraction of the global general field
population at low redshift, the exact value of such fraction depending
of course on the adopted definition of what is superdense.
%There are however two main differences with previous works.....
It remains however to be understood why the properties of galaxies 
selected applying Trujillo's criteria on the SDSS are so much different
from those found with Taylor's and our selection
(see also discussion in Taylor et al. 2010 about the comparison
with Trujillo et al. 2009).

In a separate work, Trujillo et al. (2011) investigated the issue 
of the age dependence on galaxy size using a sample of visually classified
ellipticals from the SDSS, and an HST GOODS sample at higher-z.
They found no trend of size with MW age, at fixed mass,
both in the SDSS and in the $z \sim 1$ sample. 
%The stellar age they
%showed results for is the MW age, and 
%such result is to be expected, given 
This agrees with
the consistency between the MW age of SDGs and all galaxies that
we also find in Fig.~\ref{fig:mcarlo}. However, these authors also 
mention that their result
does not change if they used LW ages. Interestingly, they do find a
variation of size with MW age at fixed {\it dynamical}
mass, in the direction of older galaxies being more compact, as we
find.

It should be noted that they selected visually morphologically
classified ellipticals (no S0s included), while the bulk of the age
selection effects should show up when including passive galaxies with
a disk component, especially S0s that are a very important component
of the passive/early-type population, representing $\sim 50$\% and
60\% of the total early-type massive galaxy population in the general
field and in clusters at low-z, respectively (Calvi et al. 2012).
Disky galaxies are common among high-z superdense massive
galaxies (van Dokkum et al. 2008, Cimatti et al. 2008,
van der Wel 2011, and other high-z references - see also
Chevance et al. 2012 for the possibility that structurally
high-z massive compact galaxies are unlike those of any local galaxy sample), 
and they dominate the SDG populations at low-z both in
the field and in clusters, as shown in Table~2.  
It is also important to stress that the size-age trend at fixed mass
is much more pronounced in clusters than in the general field, as we discussed
in \S5, therefore it is harder to detect in the field than in clusters.

Finally there are a few works in the literature investigating the
environmental dependence of the relation between galaxy size and mass,
none of them in the local Universe.  The lowest redshift study is the
one of Maltby et al. (2010)  at $z \sim 0.17$, 
who investigated the stellar-mass-size
relation for ellipticals, lenticulars and spirals, separately, in the
A901/2 STAGES field.  Using major axis effective
radii and COMBO-17 masses, they analyzed a cluster, a cluster core and
a field sample, the latter selected in the cluster area on the sky
in redshift intervals that avoid
the cluster. For ellipticals, lenticulars and high-mass spirals they
found no evidence for an environmental dependence, while for
intermediate/low mass spirals, field galaxies are larger at $2\sigma$
than in the cluster. 
An analysis of the mass-size relation for each morphological
type is beyond the scope of this paper, but in future works it would be
interesting to assess how the environmental dependence of the size-mass
relation is related to the variation of morphological mix with
environment.

Going to higher redshifts, apparently contrasting results have been found.
Raichoor et al. (2012) found that visually classified
early-type galaxies, mostly at $M_{\star} < 2 \times 10^{11} M_{\odot}$,
are more compact in the Lynx supercluster at $z=1.3$ 
than in the field
at similar redshifts, with a 35-45\% of the cluster population
being SDGs according to our definition.
The evolution observed for the cluster mass-size relation is consistent with 
the hypothesis that larger size, bulge-dominated spirals transform
into earlier types and  provoke the shift of the early-type
mass-size relation with redshift (Mei et al. 2012).
In contrast, Rettura et al. (2010) found comparable mass versus size
relations in red sequence galaxies in a cluster and in the field at $z
\sim 1.2$. Moreover, quiescent (by a combination of colors) galaxies
in a proto-cluster at $z \sim 1.6$ have been found to have {\it larger}
average effective sizes compared to quiescent field galaxies at fixed
mass, resulting in a deficit of compact galaxies 
and suggesting an accelerated evolution in
clusters compared to the field (Papovich et al. 2012).
Even more recently, Huertas-Company et al. (2013) do not detect
an environmental dependence of the galaxy mass-size relation
and the size growth in the field-to-group halo mass
range ($<10^{14} M_{\odot}$) in COSMOS and SDSS.
The environmental dependence of the mass-size relation at high-z is 
clearly still an open issue.

\section{Summary}

In this work we have analyzed the Padova Millennium Galaxy and Group Catalogue
to search for SDGs in the local Universe, study their properties and
investigate how their fraction and characteristics change with environment,
using as comparison the WINGS sample of cluster galaxies.
Our main results can be summarized as follows.

SDG galaxies, defined to have densities $\geq 3 \times 10^9 \, 
\rm M_{\odot} \, kpc^{-2}$, represent 4.4\% of all galaxies
at masses $3 \times 10^{10} M_{\odot} < M_{\star} < 4 
\times 10^{11} M_{\odot}$. This corresponds to a number density of
$4.3 \times 10^{-4} \, \rm h^{3} \, Mpc^{-3}$.
If we used literature radii and mass estimates instead of our own,
we would get similar numbers.
The SDG radii and mass densities distributions resemble those of
high-z massive and compact galaxies.

SDGs tend to be flattened, with an average intermediate Sersic
index ($<n>=2.8$), consistent with the fact that 70\% are S0s.
They are generally red and passive, with a median LW and MW ages
of 5.4 and 9.2 Gyr, respectively.
Their velocity dispersions and dynamical mass estimates are
consistent with them being exceptionally dense
compared to the general galaxy population at similar masses.

The incidence of SDGs and their luminosity-weighted ages depend
strongly on environment. The SDG fraction is 3 times higher 
and the median age 4 Gyr older in clusters
above $500 \rm \, km \, s^{-1}$ than in the general field.
Other SDG properties, such as morphological distribution
and median Sersic index, are very similar for cluster and field SDGs. 
We find no evidence for a trend with local galaxy density,
neither a change with global environment going from
isolated galaxies to binary systems to groups.

We study how the mass-size relation changes with the stellar age
of the galaxy and with environment. We find that 
the mass-size relation shifts to smaller radii for older
luminosity-weighted galaxies. This effect is mild in the field,
and much stronger in clusters.

Moreover, cluster SDGs with old luminosity-weighted ages represent
a significant fraction of all old SDGs at low-z. Unfortunately
we can only place a hard lower limit to the fraction of SDGs
that are in clusters, which is 17\%.
These old SDGs 
have sufficiently old stellar populations to be the descendants
of the high-z galaxy population assuming they have not resumed a star
formation activity at later times.

Based on these results and supported by the expectations of
hierarchical simulations that predict that 60\% of galaxies that are
already massive ($>10^{11} M_{\odot}$) and passive (SFR $< 1 M_{\odot}
yr^{-1}$) at $z \sim 2$ end up in clusters at $z=0$, we argue
that, in order to assess the amount of size
evolution of individual galaxies,
it is more appropriate  to compare the high-z data 
with {\it old, cluster} SDGs at low-z than with the general
field population of all ages.

An important fraction of the high-z galaxies is expected to be the
progenitors of today's cluster BCGs, which in fact have undergone
an evolution in size and mass that is much stronger than that of
all other galaxies. For these galaxies, the evolution has 
been a factor 4 in size and 2 in mass between $z \sim 0.7$
and $z \sim 0.05$ (Valentinuzzi et al. 2010b). These
are of course the most massive galaxies locally, usually above
the mass limit we adopt here ($M_{\star} > 4 \times 10^{11} M_{\odot}$,
see Fig.~5 in Valentinuzzi et al. 2010). This is consistent
with a mass-dependent size evolution found by other studies
(Huertas-Company et al. 2012, Ryan et al. 2012), which we speculate
might be due to BCGs.

When taking age and environmental effects into account, we find that
the average evolution in size of individual, non-BCG galaxies
of masses $3 \times 10^{10} M_{\odot} < M_{\star} < 4 
\times 10^{11} M_{\odot}$ is very mild, a factor $\sim 1.6$.

One of our most intriguing results is the dependence of the LW and MW
ages on galaxy mass, size and environment.  The LW age, that can be
considered a sort of ``quenching age'' approximately equal to the
time elapsed since galaxies stopped forming stars, is significantly influenced
by mass, size and environment.  Both in clusters and in the field, at
a given size more massive galaxies have older LW ages and, at a given
mass, smaller galaxies have older LW ages. At fixed mass and size,
cluster galaxies have older LW ages than field galaxies.
The existence of a link between mass, structure and stellar age
was already discussed in D'Onofrio et al. (2011) based on different
arguments.

The MW age, that represents the ``true'' average age of stars in
galaxies and is more related to the epoch when galaxies formed the
bulk of their stars, depends strongly on galaxy mass and environment,
and less strongly on galaxy size. These findings may help guiding models of
galaxy formation and evolution to identify the mechanisms determining the
birth epoch of most stars and the quenching epoch 
of star formation in galaxies.

%% Authors who wish to have the most important objects in their paper
%% linked in the electronic edition to a data center may do so by tagging
%% their objects with \objectname{} or \object{}.  Each macro takes the
%% object name as its required argument. The optional, square-bracket 
%% argument should be used in cases where the data center identification
%% differs from what is to be printed in the paper.  The text appearing 
%% in curly braces is what will appear in print in the published paper. 
%% If the object name is recognized by the data centers, it will be linked
%% in the electronic edition to the object data available at the data centers  
%%
%% Note that for sources with brackets in their names, e.g. [WEG2004] 14h-090,
%% the brackets must be escaped with backslashes when used in the first
%% square-bracket argument, for instance, \object[\[WEG2004\] 14h-090]{90}).
%%  Otherwise, LaTeX will issue an error. 

\acknowledgments
We would like to thank the referee, Marc Huertas-Company, for his
very constructive and careful report.
We gratefully acknowledge useful discussions with 
Michele Cappellari, Alvio Renzini and Nacho Trujillo, and we are very
grateful to Paolo Cassata for providing unpublished results for 
his high-z sample.
We acknowledge financial support from PRIN-MIUR2009 and ASI contracts
I/016/07/0 and I/009/10/0.
GDL acknowledges  financial support from the European Research Council under
the  European Community's  Seventh Framework  Programme  (FP7/2007-2013)/ERC
grant agreement n. 202781.

%% To help institutions obtain information on the effectiveness of their
%% telescopes, the AAS Journals has created a group of keywords for telescope
%% facilities. A common set of keywords will make these types of searches
%% significantly easier and more a\\urate. In addition, they will also be
%% useful in linking papers together which utilize the same telescopes
%% within the framework of the National Virtual Observatory.
%% See the AASTeX Web site at http://www.journals.uchicago.edu/AAS/AASTeX
%% for information on obtaining the facility keywords.

%% After the acknowledgments section, use the following syntax and the
%% \facility{} macro to list the keywords of facilities used in the research
%% for the paper.  Each keyword will be checked against the master list during
%% copy editing.  Individual instruments or configurations can be provided 
%% in parentheses, after the keyword, but they will not be verified.

%{\it Facilities:} \facility{Nickel}, \facility{HST (STIS)}, \facility{CXO (ASIS)}.

%% Appendix material should be preceded with a single \appendix command.
%% There should be a \section command for each appendix. Mark appendix
%% subsections with the same markup you use in the main body of the paper.

%% Each Appendix (indicated with \section) will be lettered A, B, C, etc.
%% The equation counter will reset when it encounters the \appendix
%% command and will number appendix equations (A1), (A2), etc.

\appendix

\section{Galaxy size, axial ratio and Sersic index catalog}

We release the catalog of sizes, axial ratios and Sersic indices
measured with GASPHOT for the PM2GC mass-limited galaxy sample used in
this paper. Table~A1 shows a subset of the catalog, available
in full as online material at the journal and at the CDS.
The catalog of positions, galaxy stellar masses, redshifts, environment
and other galaxy properties
can be found in Calvi et al. (2011). The morphological catalog
is released with Calvi et al. (2012, submitted),
and upon request to the author before publication.

GIM2D radii, axial ratios and Sersic indices used in this
paper are given in Allen et al. (2006) and can be found at the
Millennium Galaxy Catalogue web page http://www.eso.org/~jliske/mgc/,
file gim2d\_sersic0.92. Images are also released by the MGC,
following the instructions at the same web page.

% Table for CDS is table_paper.dat + README_table

\begin{table}
\begin{center}
\begin{tabular}{lccc}
%\hline
%\hline
\hline
ID & $R_e$ (pixel) & n & b/a \\
\hline
  61514 &  9.553  &    7.797 &   0.488 \\  
  16076 &  13.133 &    0.854 &   0.678  \\
  55832 &  9.457  &    3.295 &   0.532  \\
  22663 &  9.844  &    4.566 &   0.947  \\
  43503 &  8.667  &    3.111 &   0.779  \\
  18454 &  5.676  &    3.751 &   0.332  \\
  37797 &  10.338 &    1.242 &   0.576  \\
  18855 &  5.477  &    3.544 &   0.426  \\
  12239 &  7.964  &    3.887 &   0.592  \\
  19977 &  6.088  &    3.251 &   0.618  \\
  24151 &  5.372  &    2.096 &   0.413  \\
  ..... &  .....  &    ..... &   .....  \\
\hline 
\end{tabular}
\caption{Surface photometry parameters computed
with GASPHOT for the PM2GC mass-limited sample.
Columns: (1) PM2GC ID number as in Calvi et al. (2011); (2)
Galaxy major axis radius in pixel. The pixel size is 0.333 arcsec;
(3) Sersic index; (4) Axial ratio b/a. The full catalog is available
as online material at the journal and at the CDS.
\label{tab:catalog}}
\end{center}
\end{table}

\clearpage

\clearpage

%% If the table is more than one page long, the width of the table can vary
%% from page to page when the default \tablewidth is used, as below.  The
%% individual table widths for each page will be written to the log file; a
%% maximum tablewidth for the table can be computed from these values.
%% The \tablewidth argument can then be reset and the file reprocessed, so
%% that the table is of uniform width throughout. Try getting the widths
%% from the log file and changing the \tablewidth parameter to see how
%% adjusting this value affects table formatting.

%% The \dataset{} macro has also been applied to a few of the objects to
%% show how many observations can be tagged in a table.

\clearpage

%% The following command ends your manuscript. LaTeX will ignore any text
%% that appears after it.

\end{document}